# Measuring Value in Healthcare

*Christopher Gardner*
*i*Value LLC

# Measuring Value in Healthcare

*Christopher Gardner*
*i*Value LLC

Abstract
*A statistical description and model of individual healthcare expenditures in the US has been developed for measuring value in healthcare. We find evidence that healthcare expenditures are quantifiable as an infusion-diffusion process, which can be thought of intuitively as a steady change in the intensity of treatment superimposed on a random process reflecting variations in the efficiency and effectiveness of treatment. The arithmetic mean represents the net average annual cost of healthcare; and when multiplied by the arithmetic standard deviation, which represents the effective risk, the result is a measure of healthcare cost control. Policymakers, providers, payors, or patients that decrease these parameters are generating value in healthcare. The model has an average absolute prediction error of approximately 10-12% across the range of expenditures which spans 6 orders of magnitude over a nearly 10-year period. For the top 1% of the population with the largest expenditures, representing 20%-30% of total spending on healthcare, a power-law relationship emerges. This relationship also applies to the most expensive medical conditions in the US. A fundamental connection between healthcare expenditures and mathematical finance is found by showing that the process healthcare expenditures follow is similar to a widely used model for managing financial assets, leading to the conclusion that a combination of these two fields may yield useful results.*

*Introduction*
The history of the modeling of healthcare expenditures is extensive and previous researchers have done much to build on. Some have focused on healthcare expenditures explicitly, while others have focused on related areas such as length of stay or time in treatment. However, "few methods have been proposed for modeling the entire distribution and for prediction" as of 2006.[1] We report here progress made towards this end.

A list of the difficulties encountered can be traced to the nature of the distribution of healthcare expenditures in the population. It exhibits high asymmetry (i.e., skewness), the influence of extreme values (i.e., a heavy right tail), and a large subpopulation with zero or token amounts of expenditure (i.e., bimodality). These characteristics have led to problems such as fitting the upper and lower tail simultaneously,[2,3,4] the lack of computer resources for analysis in the past,[5] reconciling the mathematics with a generating mechanism,[6] and not ruling out competing hypotheses.[7]

The mathematical model of healthcare expenditure we have developed can be thought of intuitively as a deterministic process of increasingly intensive (and specialized) treatment superimposed on a random process reflecting variations in the efficiency and effectiveness of treatment. The deterministic process is an infusion (i.e., intensification) while the random process is a diffusion. Applying the model to the healthcare system requires no knowledge of the detailed processes that generate healthcare expenditures.

While the model describes the cost of illness in the healthcare system, at its most fundamental level it reduces to the survival time of bacteria in a disinfectant, with the stipulation that the concentration of the disinfectant is allowed to increase. The analogy is between cost and time. As this suggests, development of the model has been influenced by the work of many researchers.[8, 9, 10, 11] There is extensive literature on similar constructions across a wide-range of fields.[12]

The chief attributes that give us confidence in our result are the model's accuracy, plausibility, tractability, ability to withstand multiple independent tests, and the useful implications that arise.

This paper is organized according to a five-step approach: 1) describe the data statistically, 2) develop an intuitive and plausible process for generating healthcare expenditures and derive a mathematical model from it, 3) compute statistics on the goodness-of-fit to the data, 4) determine if the null hypothesis that the model generated the data can be shown to be false,[13] and finally 5) summarize the predicted behaviors and implications of the findings.[14]

*1. Statistical Description of Data*
The source of the healthcare expenditure data is the Medical Expenditure Panel Survey (MEPS) Household Component for 1996-2004. In the interest of space, the results for 2000, 2002, and 2004 will be primarily used for illustration.

MEPS provides annual estimates of healthcare expenditures for the U.S. civilian population. It is sponsored by the Agency for Healthcare Research and Quality (AHRQ) and the National Center for Health Statistics (NCHS). MEPS collects data through interviews with a family member who reports on the healthcare expenditures of the entire family. Two calendar years of medical expenditure data are collected for each household.[15] Approximately 10,000 households were surveyed in 2000.

The population is a mixture of individuals with zero healthcare expenditure (i.e., loosely, those who did not become ill) and individuals with non-zero expenditure (i.e., loosely, those who became ill). The total sample sizes are n=23,839 individuals for 2000; n=37,418 for 2002; and n=32,737 for 2004. Our focus is on the individuals who have a level of healthcare expenditure $\geq$ \$1, and the rest of this paper analyzes this subpopulation. The sample sizes for this group are n=19,380 individuals for 2000; n=30,728 for 2002; and n=26,507 for 2004.

As a rule of thumb, 100 unweighted cases are needed to support national estimates made in official MEPS research reports.[16] In a departure from this policy we will include every data point and use standard techniques for identifying outliers in the analysis. Goodness-of-fit tests and consistency over the 9 year period studied are relied on to give confidence in the results.

Expenditures in MEPS are defined as the sum of direct payments for healthcare provided during the year, including out-of-pocket payments and payments by private insurance, Medicare, Medicaid, and other sources. Payments for over the counter drugs, alternative care services, and phone contacts with medical providers are not included.[17] All estimates from MEPS are weighted with the weights provided in the public use files to adjust for sample design and survey nonresponse. The MEPS data is provided in the form of a distribution of population frequency versus healthcare expenditure, $f(x)$, where x is the annual individual healthcare expenditure. A listing of the logarithmically binned data points is shown in Appendix A.

We use a graphical approach as a guide to find the basic structure of the healthcare expenditure data. The advantage of this approach is that it applies no a priori restrictions on the nature of the data. An in-depth analysis is then conducted which requires limiting the set of models considered

We begin by plotting the observed frequency function of the healthcare expenditure per person. This shows the number of individuals, $f(x)$, as a function of annual healthcare expenditure per person, x. The data has been binned to be equally spaced on a logarithmic scale with each bin sized a factor of 2 wide $2^i \leq x < 2^{i+1}$ and the midpoint approximation used. A range of annual individual healthcare expenditures from \$1 to \$1,048,578 is thus comprised of 20 bins.[18] Probabilities are estimated by taking the frequency of observations at these intervals.

Plotting the frequency function on a log-linear scale $\{\log_{10}[x], f(\log_{10}[x])\}$ reveals a normal distribution (Figure 1). While the fit is not perfect, especially in the peak and upper tail, it is certainly suggestive of the classic lognormal distribution for x.

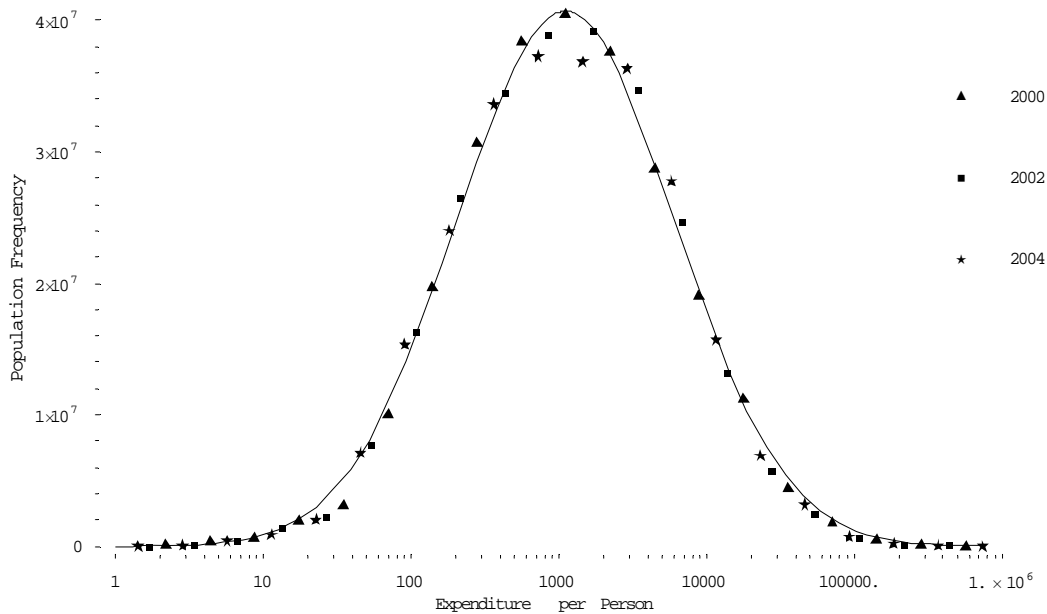

**Figure 1. Histogram of Healthcare Expenditure per Person in the US (Normalized Populations, 2004 Dollars)**

To aid comparison, the 2000 and 2002 data has been re-scaled to 2004 levels by normalizing the total US population and total US healthcare expenditure (2004 data is unchanged). As shown in Figure 1, the data for 2000, 2002, and 2004 collapse onto a similar curve,

$$p(\log_{10}[x]) = 4.07 \times 10^7 \, e^{-0.91 \, (\log_{10}[x] - 3.05)^2}$$

so the distributions can be considered to have the same functional form and are stationary after de-trending.

Let's look at the 2004 original unscaled data which is fairly representative of 2000 and 2002 (Figure 2). We have included the standard errors in the data as reported by MEPS. The peak is somewhat outside the estimated range of error.

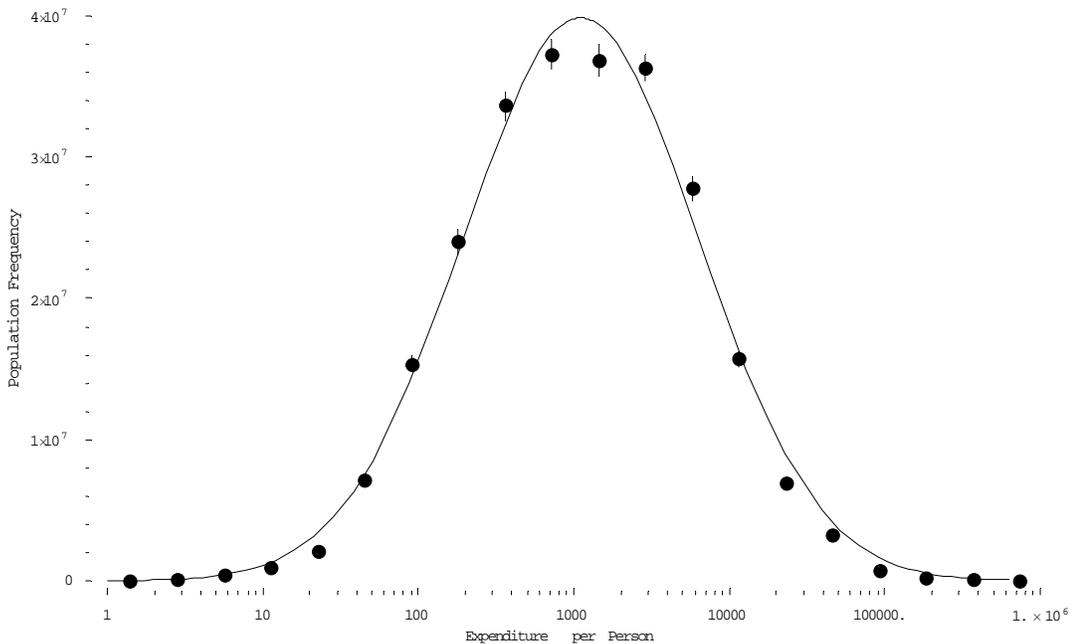

**Figure 2. Population Frequency vs. Healthcare Expenditure per Person in the US for 2004 with Standard Errors**

To zoom in on the upper tail and visually assess the quality of fit to the lognormal there, we use a technique called a Zipf plot (Figure 3).[19] The Zipf plot is a $\log_{10}$-$\log_{10}$ plot of individual healthcare expenditure against rank in the sample population based on the relation

$$\frac{i}{N} \to 1 - CDF(x_i)$$

where the observations $\{x_1, \ldots, x_i, \ldots, x_N\}$ are ordered from largest to smallest so that the index $i$ is the rank of $x_i$. The arrow means the distributions converge for large N and CDF is the abbreviation for Cumulative Distribution Function. For the lognormal, this relation can be rewritten as

$$\log_{10}[x_i] = \sqrt{2} \, \sigma \, \text{InverseErf}\left[2\left(1 - \frac{10^{\log_{10}[i]}}{N}\right) - 1\right] + \mu$$

Version 1.2                                                                                                                              4

where $\mu$ is the mean and $\sigma$ the standard deviation. It shows the imperfections in the fit, making it clear that the lognormal suffers from overestimation at the upper end of the tail, starting in the vicinity of the point where x ≳ $10,000 (Figure 3). This pattern reoccurs with 2002 and 2000 data.

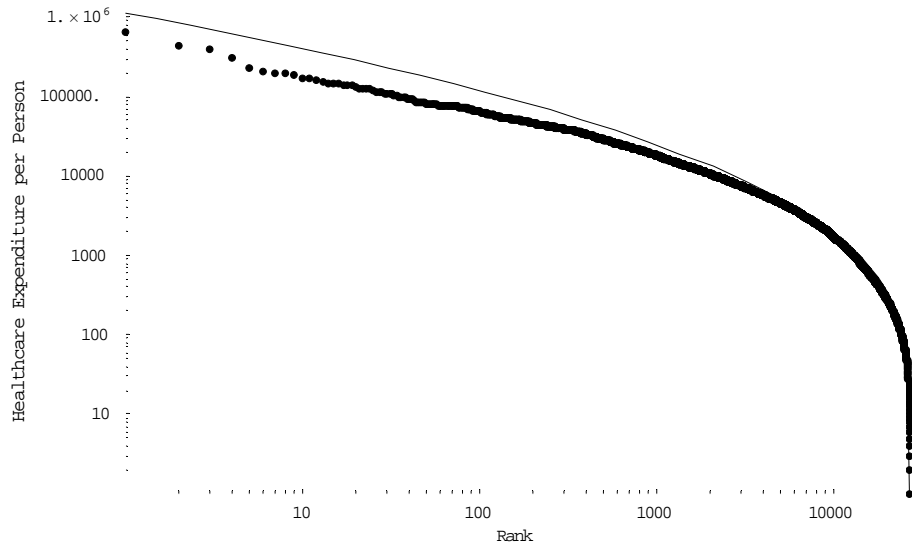

**Figure 3. Healthcare Expenditure per Person vs. Rank in the Sample Population for 2004 Compared to Curve Predicted by a Lognormal Fit**

Interestingly, the data points in the upper tail are lower compared to the prediction of the lognormal, rather than higher as would be widely believed. This gap can be considered large because the scale of the Zipf plot is logarithmic, and despite the small sample size as the top rank ($\log_{10}[1]=0$) is approached, the gap is also statistically significant. The probability that none of the other observations would exceed that of the top expenditure observed is .027 which is below conventional standards for significance. The unscaled data for 2002 and 2000 gives similar results. This raises the question: What statistical model can close the gap in the fit to the data? A clue to the answer is provided by the systematic deviation in the fit. It suggests a variant of the lognormal where the parameters are dynamic.

With this as a backdrop and motivation, we will turn our attention to developing a plausible model that accounts for the features of the MEPS healthcare expenditure data.

## *2. Mathematical Model*
We will develop a 3 and 4-parameter mathematical model.

### *3-Parameter Model*
Consider the population of individuals whose medical treatment is the responsibility of the U.S. healthcare system. Observe these individuals over a period of time sufficiently long to generate stationary statistics with suitable de-trending. This means the probability distribution does not appreciably change during this period, and as a result parameters such as the mean and variance also do not change.

Let $n_{pop}$ be the population of individuals who have a level of healthcare expenditure ≥ $1. Rank these individuals in order from lowest to highest expenditure. The discrete healthcare expenditure of each individual, which concludes in a discharge from the healthcare system, is defined as:

```
x₁   = healthcare expenditure of the first individual
x₂   = "                        "   second individual
 •
 •
 •
xⱼ₋₁ = "                        "   j - 1 individual
xⱼ   = "                        "   j   individual
 •
 •
 •
x_n_pop = healthcare expenditure of the last individual
```

Discharge is the event that marks the point in time when no more expenditure occurs for the individual for the rest of the period examined; discharge can be due to cure, death, or dropout. If an individual has intermittent episodes of illness, we combine the expenditures as if they were incurred in a single continuous episode.

We now make three assumptions. First, the discharge of an individual from the healthcare system is independent of the discharge of any other individual in the system. Second, the incremental expenditure between successive discharges from the healthcare system is proportional to the prior expenditure.[20]

$$\Delta x_j \propto x_{j-1} \text{ where } \Delta x_j = x_j - x_{j-1}$$

This is equivalent to saying that the more difficult the illness is to treat the more rapidly the expenditure increases. Third, the coupling is a combination of a deterministic and a probabilistic term [21]

$$x_j - x_{j-1} = (\delta \Delta x_j + \xi_j) x_{j-1} \text{ for } j = 1, \ldots, k, \ldots, n_{pop}$$

The deterministic term is a non-random variable that grows at the rate $\delta \Delta x_j$ where $\delta$ can be a positive or negative real number. This term has been chosen so that when summed over all increments it has the simple form $\delta x$. It represents a steady growth in individual healthcare expenditure as the severity of illness increases.

The probabilistic term is an independent, not necessarily normal random variable, $\xi_j$, with a mean and non-zero variance. The variance represents the individual patient's characteristics that shape the response to treatment, the healthcare system's characteristics that shape the efficiency and effectiveness of treatment, the illness's characteristics that shape its frequency and severity, and similar fluctuating factors that affect the discharge outcome. A picture of this is given in Figure 4.

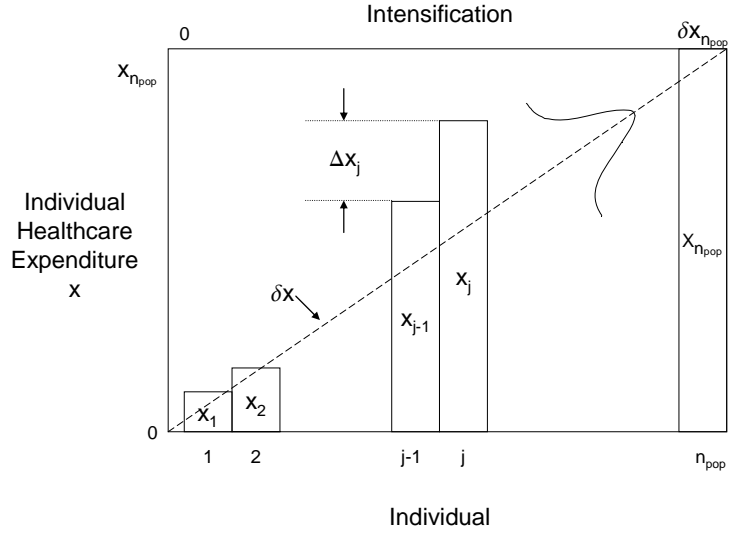

**Figure 4. Picture of 3 Parameter Mathematical Model ($\mu$, $\sigma$, $\delta$)**
*Healthcare expenditure for each individual in the population is shown against the x axis and y axis. The contribution of the probabilistic term is indicated by the bell-shaped curve, and that of the deterministic term by the dotted line which is shown against the top axis with individual healthcare expenditure acting as the x axis*

Rearranging and summing the previous equation gives

$$\sum_{j=1}^{k} \frac{x_j - x_{j-1}}{x_{j-1}} = \sum_{j=1}^{k} \delta \, \Delta x_j + \sum_{j=1}^{k} \xi_j$$

Let $n_{pop} \geq k \gg 1$. By the central limit theorem, the sum of $\xi_j$ is normally distributed with mean $\mu$ and variance $\sigma^2$, which is represented by $N[\mu, \sigma^2]$. Replacing the remaining sums with integrals gives

$$\int_{x_0}^{x_k} \frac{1}{x} \, dx = \int_{x_0}^{x_k} \delta \, dx + N_{x_0}^{x_k}[\mu, \sigma^2]$$

Since k is large, $x_k$ can be replaced by the continuous variable x. We can then write

$$\ln[x] - \ln[x_0] = \delta \, (x - x_0) + N[\mu, \sigma^2]$$
$$\ln[x] = N[\ln[x_0] + \mu + \delta \, (x - x_0), \sigma^2]$$

So for the 3-parameter infusion model the numbers $\ln[x]$ follow a normal distribution,

$$p(\ln[x]) = \frac{C}{\sqrt{2 \pi \sigma^2}} \exp\left\{ -\frac{(\ln[x] - \ln[x_0] - \mu - \delta \, (x - x_0))^2}{2 \sigma^2} \right\}$$

where C is a constant chosen such that

$$\int_0^{x_{max}} p(\ln[x]) \, dx = 1$$

This distribution has mean

$$\ln[x_0] + \mu + \delta \, (\bar{x} - x_0)$$

and variance $\sigma^2$ and x has a lognormal form with different values for these two quantities. It is worth observing that the normal distribution of $\ln[x]$ implies there are significant populations with large healthcare expenditures. This places great importance on the seriously ill.

The Cumulative Distribution Function (CDF) of the 3-parameter model is then
$$P(X \le x) = \int_1^x \frac{C}{\sqrt{2\pi\sigma^2}\, x} \exp\left\{-\frac{(\ln[x] - \ln[x_0] - \mu - \delta(x - x_0))^2}{2\sigma^2}\right\} dx$$
Which implies the Probability Density Function (PDF) is
$$p(x) = \frac{C}{\sqrt{2\pi\sigma^2}\, x} \exp\left\{-\frac{(\ln[x] - \ln[x_0] - \mu - \delta(x - x_0))^2}{2\sigma^2}\right\}$$
This distribution is completely specified by the parameters $\mu$, $\sigma$, $\delta$, and $x_0$. The starting point for the data is represented by $x_0 = \$1$ and therefore can generally be ignored in the calculations of interest.

An approximation to the mean and variance of x can be arrived at with some effort. Letting $\Delta\delta = \delta - 0$, we can make the Taylor Series approximation
$$\mu[\delta = \delta] \approx \mu[\delta = 0] + \Delta\delta \left(\frac{\partial \mu}{\partial \delta}\right)_{\delta=0}$$
A more useful variation of the approximation which will be seen to be important to the development of metrics and benchmarks is
$$\mu[\delta = 0] \approx \mu[\delta = \delta] - \Delta\delta \left(\frac{\partial \mu}{\partial \delta}\right)_{\delta=\delta} = \mu - \delta \left(\frac{\partial \mu}{\partial \delta}\right)_{\delta=\delta}$$
$$\approx \mu \left(1 - \frac{\Delta\mu}{\mu}\right)$$
which is evaluated at the best fit parameter $\delta=\delta$ and the sign is reversed because we are going from $\delta \to 0$ instead of going from $0 \to \delta$. We have used the explicit form for $\mu = \mu[\delta=\delta]$ to make the interpretation of the equation easier. The term
$$\frac{\Delta\mu}{\mu}$$
is the logarithmic discount/premium and its effect is to decrease/increase the average of the logarithm of healthcare expenditures leaving a *net amount*. Specifically we have
$$\frac{\Delta\mu}{\mu} = \frac{\mu[\delta=\delta] - \mu[\delta=0]}{\mu[\delta=\delta]} = \frac{\mu - \mu[\delta=0]}{\mu}$$
The term $\mu[\delta=0]$ is an estimate of what the mean of $\ln[x]$ would have to be if $\delta=0$. In other words, it is the value of the mean such that the classic lognormal model with parameters $\{\mu[\delta=0], \sigma\}$ yields an identical estimate of total healthcare expenditures as the 3-parameter model with parameters $\{\mu, \delta, \sigma\}$. This is what is meant by the logarithmic discount/premium: it is a comparison of the 2-parameter classic lognormal before and after taking into account the cumulative effect of $\delta$ on total healthcare expenditures. Letting uppercase X= total US healthcare expenditures under the classic lognormal model, $\mu[\delta=0]$ is defined mathematically as the solution to the following equation with the notation $f_2$ representing the frequency function of the 2-parameter classic lognormal model and $f_3$ the 3-parameter infusion model.
$$X(\mu[\delta=0]) \equiv \int_1^{x_{max}} x\, f_2(\mu[\delta=0], \sigma, x)\, dx = \int_1^{x_{max}} x\, f_3(\mu, \delta, \sigma, x)\, dx$$

We now have the tools to estimate the mean and variance of x. Since $\mu[\delta=0]$ is the value of the mean such that the classic lognormal model yields an identical estimate of total healthcare expenditures as the 3-parameter model, we can use the formula for the mean of x for the classic lognormal, which is called $\alpha$ in the literature.[22]

$$\bar{x} \approx e^{\mu[\delta=0] + \frac{\sigma^2}{2}}$$

$$= e^{\mu - \Delta\delta \left(\frac{\partial \mu}{\partial \delta}\right)_{\delta=\delta} + \frac{\sigma^2}{2}}$$

$$\approx e^{\mu + \frac{\sigma^2}{2}} \left(1 - \Delta\delta \left(\frac{\partial \mu}{\partial \delta}\right)_{\delta=\delta}\right)$$

$$= \alpha \left(1 - \Delta\delta \left(\frac{\partial \mu}{\partial \delta}\right)_{\delta=\delta}\right)$$

The series approximation to the exponential above is based on the empirical fact that

$$\Delta\mu = \Delta\delta \left(\frac{\partial \mu}{\partial \delta}\right)_{\delta=\delta} < 1$$

Similarly, we can use the formula for the variance of x for the classic lognormal, which is called $\beta^2$

$$\overline{x^2} - \bar{x}^2 \approx e^{2\mu[\delta=0] + 2\sigma^2} \left(e^{\sigma^2} - 1\right)$$

$$= e^{2\mu - 2\Delta\delta \left(\frac{\partial \mu}{\partial \delta}\right)_{\delta=\delta} + 2\sigma^2} \left(e^{\sigma^2} - 1\right)$$

$$\approx e^{2\mu + 2\sigma^2} \left(e^{\sigma^2} - 1\right) \left(1 - 2\Delta\delta \left(\frac{\partial \mu}{\partial \delta}\right)_{\delta=\delta}\right)$$

$$= \beta^2 \left(1 - 2\Delta\delta \left(\frac{\partial \mu}{\partial \delta}\right)_{\delta=\delta}\right)$$

The standard deviation of x is simply

$$\sqrt{\overline{x^2} - \bar{x}^2} \approx \beta \left(1 - \Delta\delta \left(\frac{\partial \mu}{\partial \delta}\right)_{\delta=\delta}\right)$$

In summary, the 3 parameter model is comprised of a random fluctuation $N[\mu, \sigma^2]$ superimposed on a growth trend $\delta x$. While both of these terms affect the level of expenditure, it is important to realize that they are separate processes when it comes to calculations. We call $\delta$ the "infusion of care" coefficient and $\sigma$ the "diffusion of care" coefficient. The mean and variance of x in the 3-parameter model can be approximated by constructing an equivalent 2-parameter model and applying the analytical solution for these quantities known for this simple form.

*4-Parameter Model*
The derivation of the 4-parameter model is similar to the 3-parameter except the infusion includes a second order term, $\delta x + \epsilon x^2$, where $\epsilon$ can be a positive or negative number. See Figure 5.

**Figure 5. Picture of 4 Parameter Mathematical Model ($\mu, \sigma, \delta, \epsilon$)**
*Healthcare expenditure for each individual in the population is shown against the x axis and y axis. The contribution of the probabilistic term is indicated by the bell-shaped curve, and that of the deterministic term by the dotted line which is shown against the top axis with individual healthcare expenditure acting as the x axis*

So for the 4-parameter model the numbers ln[x] follow a normal distribution,

$$p(\ln[x]) = \frac{C}{\sqrt{2\pi\sigma^2}} \exp\left\{-\frac{(\ln[x] - \ln[x_0] - \mu - \delta(x - x_0) - \epsilon(x^2 - x_0^2))^2}{2\sigma^2}\right\}$$

This distribution has mean

$$\ln[x_0] + \mu + \delta(\bar{x} - x_0) + \epsilon(\overline{x^2} - x_0^2)$$

and variance $\sigma^2$ and x has a lognormal form with different values for these two quantities.

The Cumulative Distribution Function (CDF) of the 4-parameter model is then

$$P(X \leq x) = \int_1^x \frac{C}{\sqrt{2\pi\sigma^2}\, x} \exp\left\{-\frac{(\ln[x] - \ln[x_0] - \mu - \delta(x - x_0) - \epsilon(x^2 - x_0^2))^2}{2\sigma^2}\right\} dx$$

Which implies the Probability Density Function (PDF) is

$$p(x) = \frac{C}{\sqrt{2\pi\sigma^2}\, x} \exp\left\{-\frac{(\ln[x] - \ln[x_0] - \mu - \delta(x - x_0) - \epsilon(x^2 - x_0^2))^2}{2\sigma^2}\right\}$$

A log-linear plot {ln[x], p(ln[x])} of the 3 and 4 parameter model compared to a best fit lognormal gives an example of how the $\delta$ and $\epsilon$ terms push weight around the tail (Figure 6).

To give an idea of the flexibility of the 4-parameter model, Appendix B has log-linear {$\log_{10}[x]$, $f(\log_{10}[x])$} plots of the distribution for several values of $\mu, \sigma, \delta,$ and $\epsilon$.

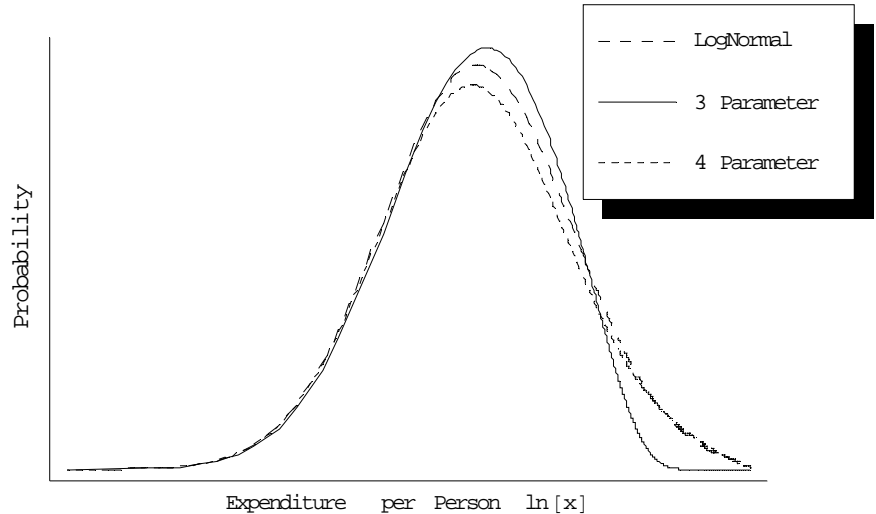
**Figure 6.  3 and 4 Parameter Model Compared to the Lognormal**

We would prefer working with the CDF of the 3 and 4-parameter distributions to improve the sample size at the extremes. A general closed-form solution, however, is elusive for the CDF, even after using powerful mathematics software programs.

A typical assumption in health services research is that medical costs are log-normally distributed.[23] The 3 and 4 parameter models reduce to a classic lognormal when there is no infusion $\delta, \epsilon = 0$. So clearly the lognormal with infusion model is compatible with this assumption. Of course many researchers have remarked that while the classic lognormal provides a better fit compared to alternative distributions, it does not fit the entire distribution well.[24]

*3. Parameter Estimates*
*Fitting the Data to the Model*
The mathematical model developed in the previous section must be converted from a probability density function to a frequency function in order to match the MEPS data. This is easily done by scaling the PDF to the total population $n_{pop}$.
    `f (x) = n_pop p (x)`
The mathematical model is nonlinear in the parameters and therefore we must apply nonlinear regression analysis to arrive at a fit to the data. This is a difficult topic of considerable complexity, but unavoidable given the nature of healthcare expenditures. We have already briefly encountered it in the statistical description of healthcare expenditures.

The standard procedure is to use a maximum likelihood estimator of the parameters. The maximum likelihood estimate of the model parameters is obtained by minimizing the $\chi^2$ merit function given by the sum of squared residuals $\Sigma_i e_i^2$. It assumes the measurement errors in the data follow a normal distribution.

To make the model as simple, linear, and symmetric as possible in the parameters, we transform the data for p(x) and x by taking their log. The 4-parameter mathematical model for f(ln[x]) written in logarithmic form is:

$$\ln[f(\ln[x])] = \ln\left[\frac{n_{pop}}{\sqrt{2\pi\sigma^2}}\right] + \ln[C_{\ln[x]}] - \frac{\left(\ln[x] - \mu - \delta\, e^{\ln[x]} - \epsilon\, e^{\ln[x^2]}\right)^2}{2\sigma^2}$$

where we have put

$$C \to C_{\ln[x]},\ x_0 \to 1,\ \text{and } x \gg x_1$$

While this is interesting, we really want the distribution of x, not ln[x]. There is some complexity here to explain. The dataset we use for f(ln[x]) is identical to that for f(x) but the equations for these quantities differ. A fit to ln[f(ln[x])] gives values for ln[x] such as the mean of the log of x, $\mu = \overline{\ln[x]}$. A fit to ln[f(x)] gives values for x in logarithmic form such as the log of the mean of x

$$\ln[\bar{x}] = \ln\left[\mathrm{Exp}\left[\mu + \frac{\sigma^2}{2}\right]\right] = \mu + \frac{\sigma^2}{2}$$

which is a relation known for the classic lognormal, and shows that for ln[f(x)] the quantity $\mu$ is subject to a transformation where [25]

$$\frac{\sigma^2}{2}$$

is added. The quadratic term in the above equation for ln[f(x)] is subject to two additive transformations. We will use this for fitting data to the regression equation written as:

$$\ln[f(x)] = \ln\left[\frac{n_{pop}}{\sqrt{2\pi}\,\sigma}\right] + \ln[C_x] - \frac{\left(\ln[x] - \mu - \sigma^2 - \delta\, e^{\ln[x]} - \epsilon\, e^{\ln[x^2]}\right)^2}{2\sigma^2} - \ln[x]$$

where

$$C_x = C_{\ln[x]}\, e^{\mu + \frac{\sigma^2}{2}}$$

The substitution

$$y = \ln[x]$$

is made and the regression is performed against the data points {ln[x], ln[f(x)]}. The regression will give estimates of the parameters $\mu$, $\sigma$, $\delta$, and $\epsilon$.

As a further aid to transparency, we will progressively examine fitting the data to 1) the classic lognormal, 2) the 3-parameter lognormal with infusion term $\delta x$, and 3) the 4-parameter lognormal with infusion term $\delta x + \epsilon x^2$. This process systematically assembles the components of our equation into an increasingly better model.

For the classic lognormal, step (1), the frequency functions for ln[x] and x can be compared to obtain a few insights:

$$f(\ln[x]) = \frac{C_{\ln[x]}\, n_{pop}}{\sqrt{2\pi\sigma^2}}\, \exp\left\{-\frac{(\ln[x] - \mu)^2}{2\sigma^2}\right\}$$

$$f(x) = \frac{C_x\, n_{pop}}{\sqrt{2\pi\sigma^2}\, x}\, \exp\left\{-\frac{(\ln[x] - \mu - \sigma^2)^2}{2\sigma^2}\right\}$$

The regression equations are:

$$\ln[f(\ln[x])] = \ln\left[\frac{n_{pop}}{\sqrt{2\pi\sigma^2}}\right] + \ln[C_{\ln[x]}] - \frac{(\ln[x] - \mu)^2}{2\sigma^2}$$

$$\ln[f(x)] = \ln\left[\frac{n_{pop}}{\sqrt{2\pi\sigma^2}}\right] + \ln[C_x] - \frac{(\ln[x] - \mu - \sigma^2)^2}{2\sigma^2} - \ln[x]$$

The relationship between $\ln[p(x)]$ and $\ln[p(\ln[x])]$ explains why a shift is observed in the constant term that deals with normalization:[26]

$$\ln[C_x] = \ln[C_{\ln[x]}] + \mu + \frac{\sigma^2}{2}$$

This relationship is corroborated by tests on artificial samples drawn from a classic lognormal.[27]

For the lognormal with infusion term $\delta x$, step (2), the frequency functions for $\ln[x]$ and $x$ are:

$$f(\ln[x]) = \frac{C_{\ln[x]} n_{pop}}{\sqrt{2\pi\sigma^2} \, x} \exp\left\{-\frac{(\ln[x] - \mu - \delta x)^2}{2\sigma^2}\right\}$$

$$f(x) = \frac{C_x n_{pop}}{\sqrt{2\pi\sigma^2} \, x} \exp\left\{-\frac{(\ln[x] - \mu - \sigma^2 - \delta x)^2}{2\sigma^2}\right\}$$

The previous discussion of transformations again applies. We have for the regression equations:

$$\ln[f(\ln[x])] = \ln\left[\frac{n_{pop}}{\sqrt{2\pi\sigma^2}}\right] + \ln[C_{\ln[x]}] - \frac{(\ln[x] - \mu - \delta\, e^{\ln[x]})^2}{2\sigma^2} - \ln[x]$$

$$\ln[f(x)] = \ln\left[\frac{n_{pop}}{\sqrt{2\pi}\,\sigma}\right] + \ln[C_x] - \frac{(\ln[x] - \mu - \sigma^2 - \delta\, e^{\ln[x]})^2}{2\sigma^2} - \ln[x]$$

For the complete 4-parameter mathematical model, step (3), the formulas have already been described.

*Outliers*

An examination of the data reveals an outlier in 2004 at x=$1. The sample size for this point is one and it is the only point in the bin $2^0 \leq x < 2^1$. The residuals show that this point has the largest studentized residual for any model in any given year: -2.62 for the classic lognormal, -3.36 for the 3-parameter infusion model, and -3.21 for the 4-parameter infusion model. Figure 7 shows a scatterplot of the 2004 data. To better reveal outliers, observations with the same value of x have been combined.

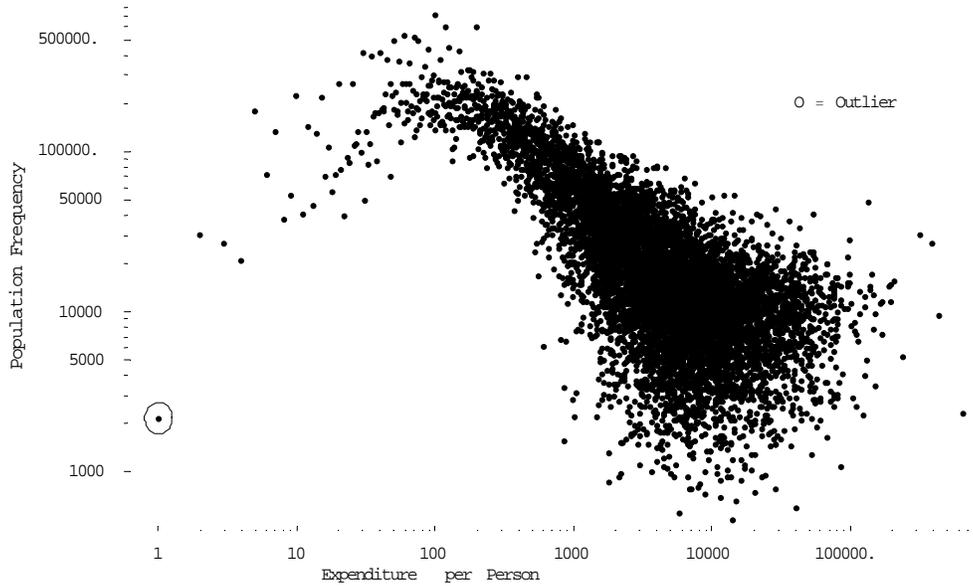

**Figure 7. Scatterplot of Healthcare Expenditure Data for 2004**

*Regression Results*
The best fit parameter estimates are shown in Figure 8. The 3 and 4-parameter models agree well with the actual value of $\mu$ which is 6.59 for 2000, 6.85 for 2002, and 6.98 for 2004. The estimate of $\mu$ significantly improves from the classic lognormal[28] to the 3 and 4-parameter models. The actual standard deviation of the data is 1.67 for 2000, 1.68 for 2002, and 1.70 for 2004. In estimating the standard deviation, the 3 parameter model appears to be about as close as the 4 parameter model.

| Classic Lognormal | 2000 | 2002 | 2004 |
|---|---|---|---|
| $\mu =$ | $6.45 \pm 0.04$ | $6.72 \pm 0.03$ | $6.85 \pm 0.06$ |
| $\sigma =$ | $1.64 \pm 0.02$ | $1.61 \pm 0.02$ | $1.58 \pm 0.03$ |
| **3 Parameter Model** | | | |
| $\mu =$ | $6.56 \pm 0.04$ | $6.78 \pm 0.05$ | $6.98 \pm 0.04$ |
| $\sigma =$ | $1.70 \pm 0.02$ | $1.63 \pm 0.02$ | $1.66 \pm 0.02$ |
| $\delta =$ | $(-2.5 \pm 0.6) \times 10^{-6}$ | $(-1.0 \pm 0.6) \times 10^{-6}$ | $(-1.6 \pm 0.3) \times 10^{-6}$ |
| **4 Parameter Model** | | | |
| $\mu =$ | $6.58 \pm 0.06$ | $6.89 \pm 0.04$ | $6.91 \pm 0.05$ |
| $\sigma =$ | $1.71 \pm 0.03$ | $1.67 \pm 0.02$ | $1.63 \pm 0.03$ |
| $\delta =$ | $(-3.9 \pm 2.9) \times 10^{-6}$ | $(-9.2 \pm 2.0) \times 10^{-6}$ | $(1.1 \pm 1.3) \times 10^{-6}$ |
| $\epsilon =$ | $(3.3 \pm 6.6) \times 10^{-12}$ | $(1.9 \pm 0.5) \times 10^{-11}$ | $(-3.2 \pm 1.5) \times 10^{-12}$ |

**Figure 8. Best Fit Parameters and Standard Errors**

A negative value is found for the coefficient of infusion $\delta$ in the 3-parameter model, which means the effect of the infusion of healthcare is to reduce expenditures. The $\delta$ term in the 3-parameter model therefore provides a quantification of the efficiencies that arise in the treatment of the increasingly ill such as from the use of specialists and hospitals. In 2004, Figure 9 shows a reduction in the population when $x \gtrsim \$6,400$ and an expansion of the population when $\$200 \lesssim x \lesssim \$6,400$ as a result of the $\delta$ term. The reduction is substantial where expenditures are highest, reaching in excess of 50%. The change in the population is calculated as the ratio of the 3-parameter model to the classic lognormal with a baseline of 1 meaning the populations are identical.

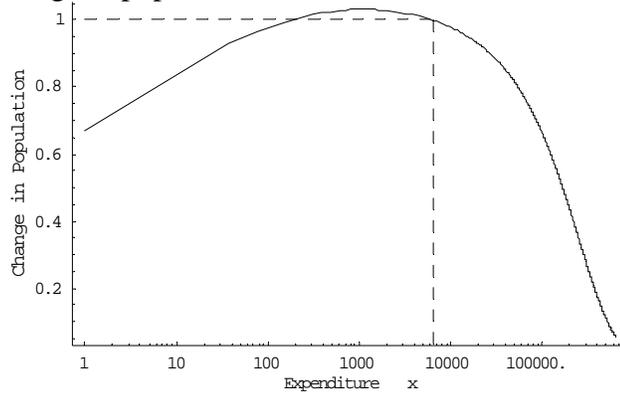

**Figure 9. Percent Change in Population Due to the Infusion Term in 2004**

The magnitude of the term $\delta x$ itself in 2004 is <1% of its maximum for annual expenditures below about $5,000. It rises to >50%, however, when annual expenditures reach $325,000.

For the 4-parameter model, the t-ratio for the coefficients of infusion $\delta$ and $\epsilon$ in 2000, as well as $\delta$ in 2004, suggests they could be estimates of zero and so the simpler 3-parameter model may be adequate for these years. The 4-parameter model reveals a deeper relationship may underlie the formation of efficiencies. Figure 10 compares the infusion equation in the 4 parameter model to the 3 parameter model.

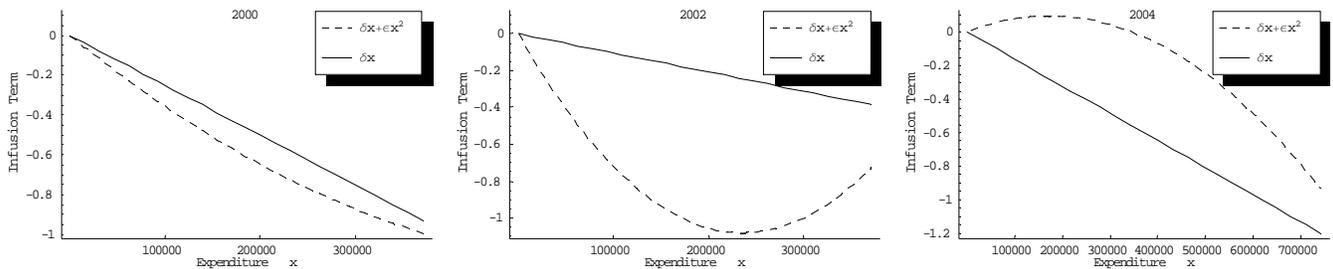

**Figure 10. Infusion Term**

The curvature of the infusion equation in the 4-parameter model is due to the offset of the linear efficiencies, $\delta x$, by quadratic inefficiencies, $\epsilon x^2$. The curvature of 2000 is noticeably less than 2002, while an entirely opposite curvature occurs in 2004. It is possible that these variations are a result of the sample size we are working with. Alternatively, we may be observing the effects of a different mix of patients, cases, and treatments from year to year in the region where healthcare expenditures are highest. A fluctuating combination of efficiencies and inefficiencies would be reflected in this mix.

The regression results for 2004 are similar to those for 2002 and 2000. We will, therefore, focus on 2004 in the remainder of this section. Plots of the data and the best fit functions for 2004 are shown in Figure 11. This visualizes the improvement in the fit from the classic lognormal to the 3 and 4-parameter models. Plots of the studentized residuals vs.

fitted values are shown in Figure 12. The residuals increase in randomness from an angled to a more cloud-like shape characteristic of uncorrelated residuals. The apparent decrease in the variance of the residuals where the population frequency is high is consistent with the larger sample size in this region. A normal probability (quantile-quantile) plot is shown in Figure 13. It gives a direct check on the assumption in a least squares fit that the residuals follow a normal distribution. The plots turn from a somewhat crooked to a fairly straight line. A straight line occurs when the fitted function is correct and the assumption of normality is appropriate. The approximate confidence bands are shown in Figure 14. They get tighter as the number of parameters increases from 2 to 3 to 4. The ANOVA table, asymptotic correlation matrix, and fit curvature table is shown in Figure 15. The ANOVA table shows the estimated variance (mean square error) drops from .13 to .05 to .04. The asymptotic correlation matrix shows that there is a fairly high correlation between $\delta$ and $\epsilon$ in the 4 parameter model, which suggests it may be over parameterized. The fit curvature table shows that the degree of nonlinearity in the fit for the 2, 3, and 4 parameter models is acceptable, and the linear approximation at the least squares estimate is valid (planar and uniform coordinates assumptions). In Figures 16 and 17 regions of reasonable parameter values are shown in pairwise plots of the parameter approximate joint 95% confidence regions, with the least squares estimate indicated by +. There is a 95% chance that the true parameter value falls within the region around the estimated value.

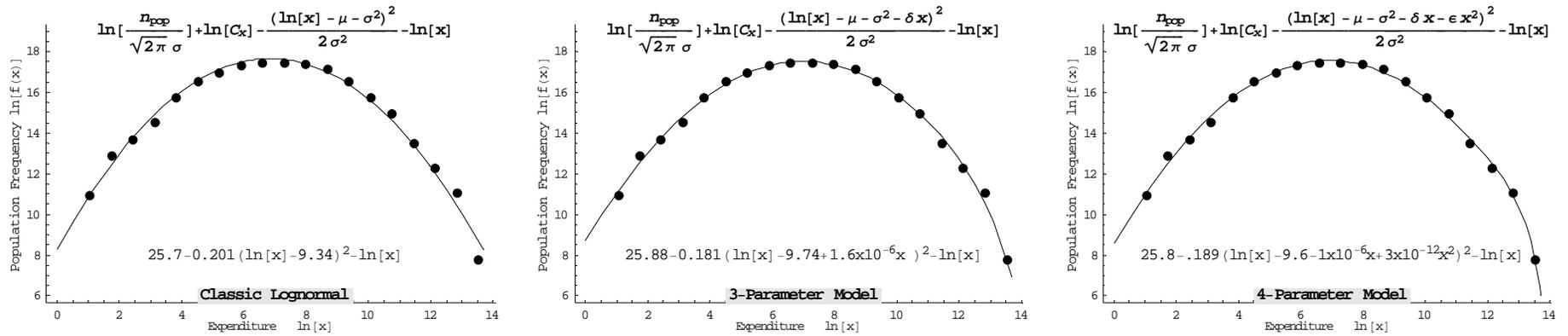

**Figure 11. Best Fits to Healthcare Expenditure Data for 2004**

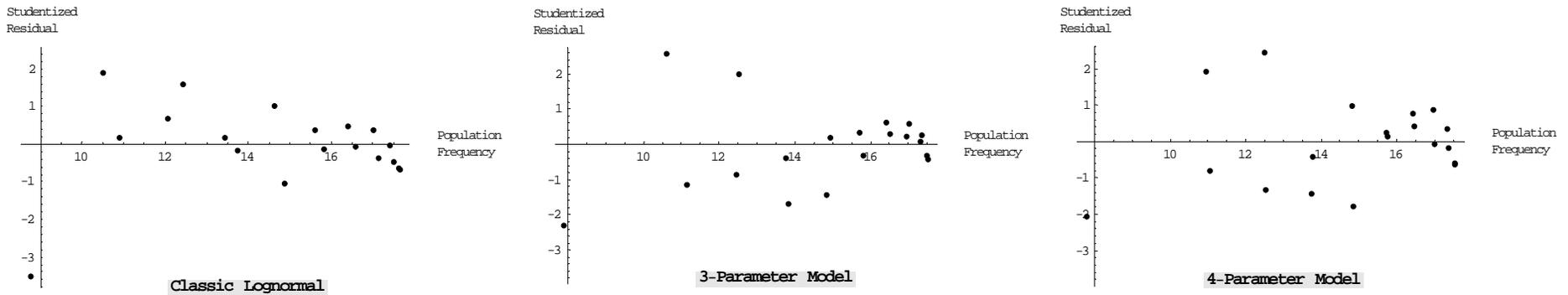

**Figure 12. Studentized Residuals vs. Fitted Values for 2004**

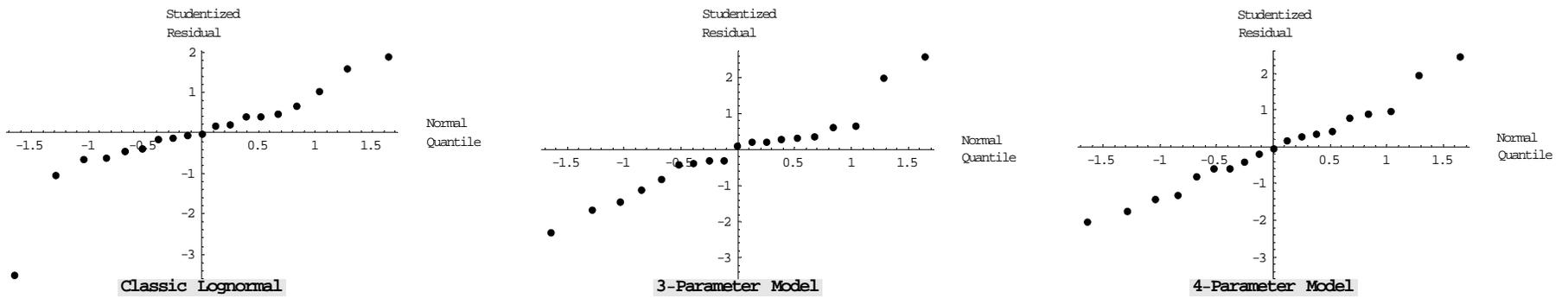

**Figure 13. Quantile-Quantile Plot of the Studentized Residuals vs. a Normal Distribution for 2004**

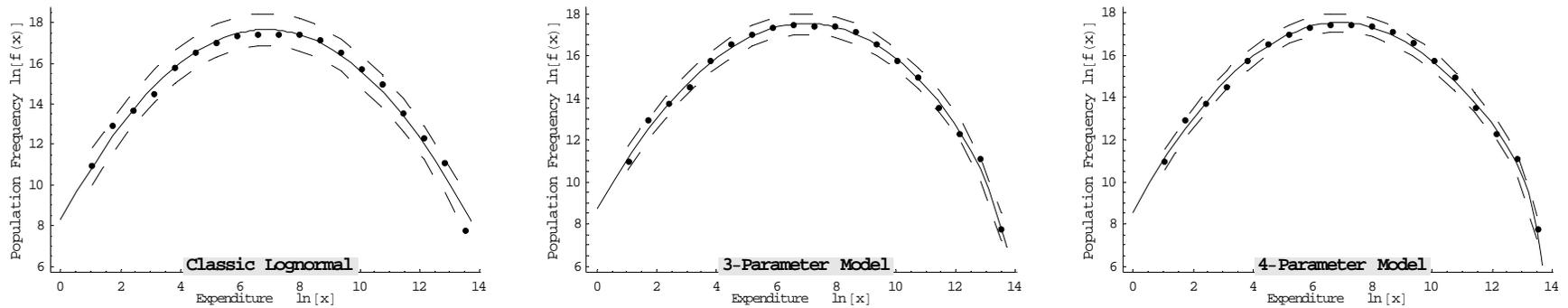

**Figure 14. 95% Confidence Interval for the Predicted Responses with Observations for 2004**

| ANOVA Table | DF | SumOfSq | MeanSq | | ANOVA Table | DF | SumOfSq | MeanSq | | | ANOVA Table | DF | SumOfSq | MeanSq | | |
|---|---|---|---|---|---|---|---|---|---|---|---|---|---|---|---|---|
| Model | 3 | 4262.59 | 1420.86 | | Model | 4 | 4263.96 | 1065.99 | | | Model | 5 | 4264.14 | 852.83 | | |
| Error | 16 | 2.08 | 0.13 | | Error | 15 | 0.72 | 0.05 | | | Error | 14 | 0.53 | 0.04 | | |
| Corrected Total | 19 | 4264.67 | | | Corrected Total | 19 | 4264.67 | | | | Corrected Total | 19 | 4264.67 | | | |
| Uncorrected Total | 18 | 136.37 | | | Uncorrected Total | 18 | 136.37 | | | | Uncorrected Total | 18 | 136.37 | | | |
| Asymptotic Correlation | $\mu$ | $\sigma$ | $\ln[C_x]$ | | Asymptotic Correlation | $\mu$ | $\sigma$ | $\ln[C_x]$ | $\delta$ | | Asymptotic Correlation | $\mu$ | $\sigma$ | $\ln[C_x]$ | $\delta$ | $\epsilon$ |
| $\mu$ | 1.00 | −0.25 | 0.52 | | $\mu$ | 1.00 | 0.26 | 0.68 | −0.56 | | $\mu$ | 1.00 | 0.55 | 0.80 | −0.72 | −0.63 |
| $\sigma$ | −0.25 | 1.00 | −0.46 | | $\sigma$ | 0.26 | 1.00 | 0.09 | −0.77 | | $\sigma$ | 0.55 | 1.00 | 0.41 | −0.68 | −0.56 |
| $\ln[C_x]$ | 0.52 | −0.46 | 1.00 | | $\ln[C_x]$ | 0.68 | 0.09 | 1.00 | −0.45 | | $\ln[C_x]$ | 0.80 | 0.41 | 1.00 | −0.68 | −0.62 |
| | | | | | $\delta$ | −0.56 | −0.77 | −0.45 | 1.00 | | $\delta$ | −0.72 | −0.68 | −0.68 | 1.00 | 0.98 |
| | | | | | | | | | | | $\epsilon$ | −0.63 | −0.56 | −0.62 | 0.98 | 1.00 |
| Fit Curvature Table | Curvature | | | | Fit Curvature Table | Curvature | | | | | Fit Curvature Table | Curvature | | | | |
| Max Intrinsic | $3.34 \times 10^{-17}$ | | | | Max Intrinsic | $2.53 \times 10^{-2}$ | | | | | Max Intrinsic | $4.93 \times 10^{-2}$ | | | | |
| Max Parameter Effects | 0.13 | | | | Max Parameter Effects | 0.12 | | | | | Max Parameter Effects | 0.13 | | | | |
| 95% ConfidenceRegion | 0.56 | | | | 95% ConfidenceRegion | 0.57 | | | | | 95% ConfidenceRegion | 0.58 | | | | |

**Classic Lognormal**        **3-Parameter Model**        **4-Parameter Model**

**Figure 15. ANOVA Table, Asymptotic Correlation Matrix, and Fit Curvature Table for 2004**

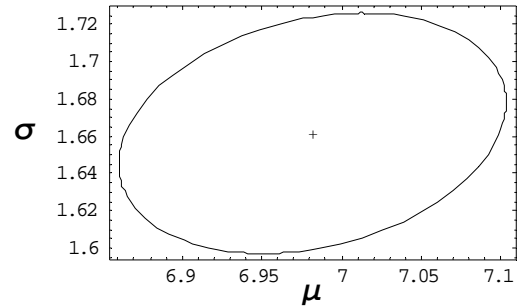

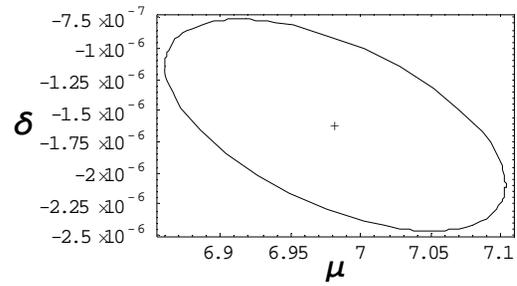
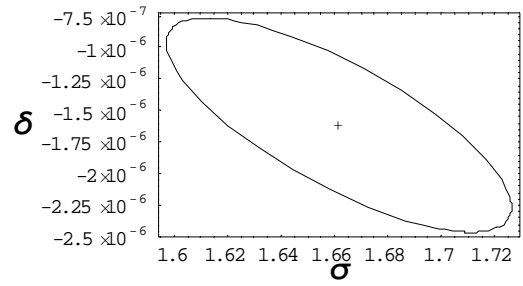

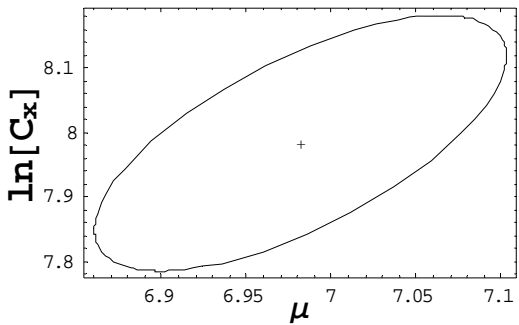
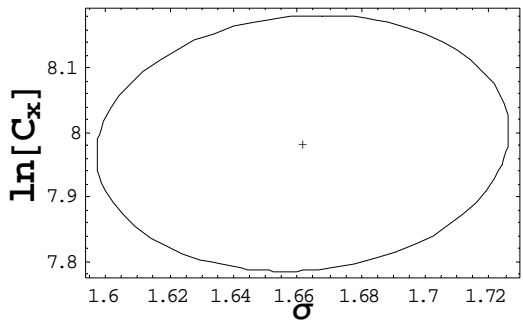
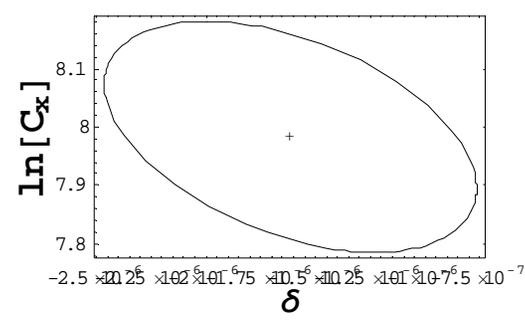

3-Parameter Model

**Figure 16. Parameter Inference Regions for 2004**

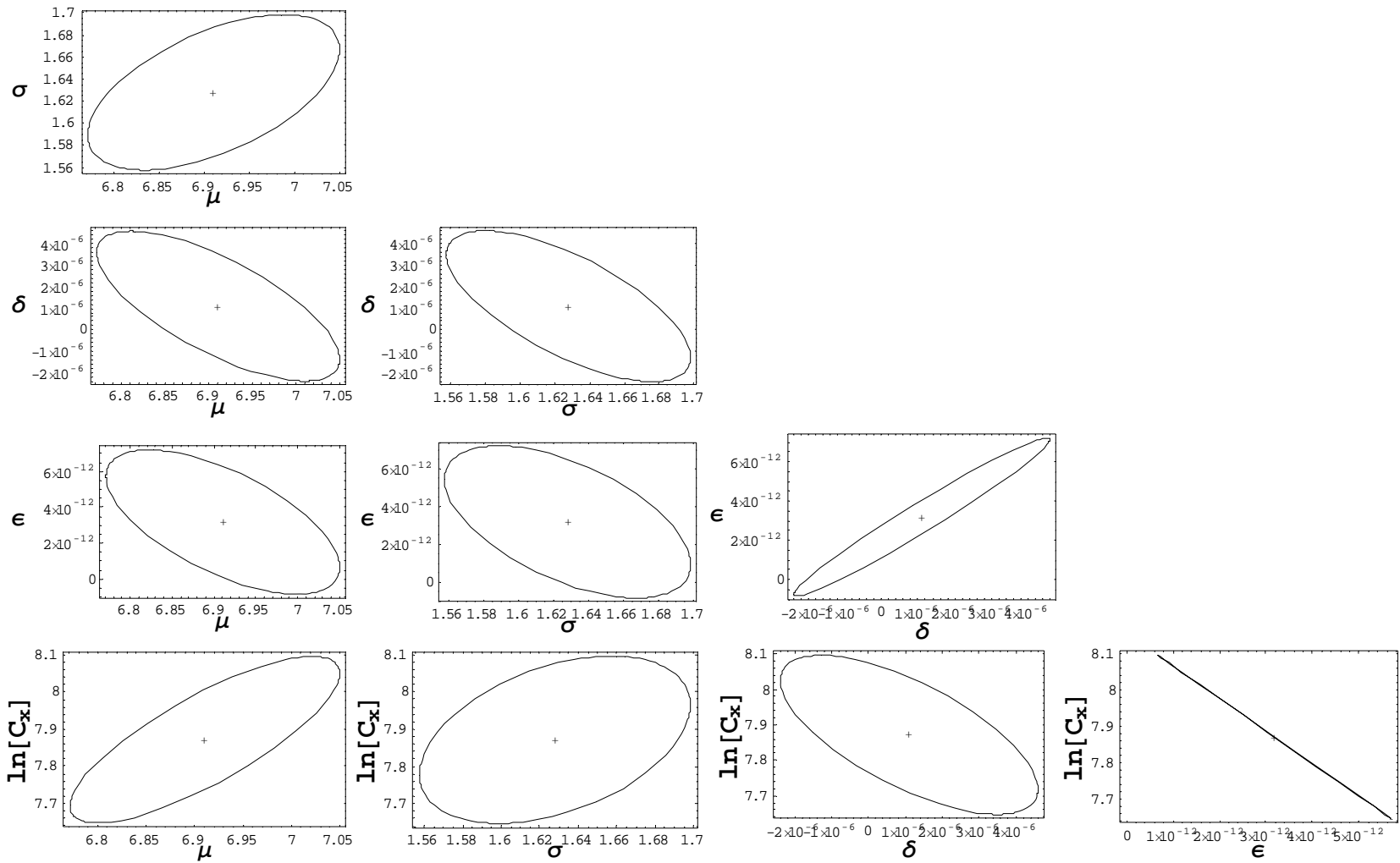

**Figure 17. Parameter Inference Regions for 2004 (Note: $\epsilon$ has sign reversed)**

*4. Goodness of Fit*

The MEPS data is not random; it is designed to be representative of the nation. For example, there is oversampling of subpopulations. The tools to assess the goodness-of-fit are therefore limited. Generally, the model that has the most explanatory power, smallest residual mean square, the most random-looking residuals, and the lowest absolute prediction error should be chosen.[29]

Based on this criterion the 4-parameter model would be preferred, but a case can be made for the 3-parameter model based on the gain in simplicity compared to the loss of accuracy. Figure 18 shows the residual mean square.

| Mean Square Error | 2000 | 2002 | 2004 | Total |
|---|---|---|---|---|
| Classic Lognormal | 0.069 | 0.045 | 0.130 | 0.244 |
| 3 Parameter Model | 0.033 | 0.041 | 0.048 | 0.122 |
| 4 Parameter Model | 0.035 | 0.020 | 0.038 | 0.093 |

**Figure 18. Mean Square Error**

Figure 19 is a table of the absolute prediction errors for the 4-parameter model. The table compares the predicted response of the 4-parameter model to the observed response using the best fit parameters for 2000, 2002, 2004. The largest deviations are concentrated in the bins where the sample size is small. The fit is strong in the far tail, which is a crucial part of the range. It also points to a possible systematic data measurement error in bin #5, assuming the model is correct. For comparison, the average absolute prediction errors are 19.3% for the classic lognormal, 12.4% for the 3-parameter model, and 10.4% for the 4-parameter.

Stringent goodness-of-fit tests such as the extra sum-of-squares test and the log likelihood ratio tests for nested models assume the data is drawn at random from the true distribution. The assessment of significance they provide, e.g., whether the difference in the log likelihoods is simply due to chance fluctuations, is biased when the sample is non-random. An experiment could be conducted where a random sample is drawn from the population.[30] A procedure for assessing the goodness-of-fit in this case is to apply three tests in succession. First, the Kolmogorov-Smirnov test applied to the classic lognormal distribution provides a baseline assessment of the goodness-of-fit. Second, the extra sum-of-squares test (or log likelihood ratio test) for nested [31] models compares the 3-parameter model to the 2-parameter classic lognormal. Third, repeating the extra sum of squares test compares the 4-parameter to the 3-parameter model.

This 3-step KS-extra sum of squares procedure [32] is similar to the systematic assembly of the components of our equation into an increasingly better model done in the previous section, except now we progressively measure the goodness-of-fit. In this way, we can determine if the observed data could have plausibly been drawn from the classic lognormal. Then we can directly compare the classic lognormal, 3-parameter, and 4-parameter models to see which, if any, is a better fit to the data. This allows us to decide if the fit is acceptable and whether a model with fewer or more parameters should be considered.

| Bin | x midpoint | SampleSize | 2000 $\frac{\text{Exp[4 ParameterPredicted]} - \text{Exp[Observed]}}{\text{Exp[Observed]}}$ | SampleSize | 2002 $\frac{\text{Exp[4 ParameterPredicted]} - \text{Exp[Observed]}}{\text{Exp[Observed]}}$ | SampleSize | 2004 $\frac{\text{Exp[4 ParameterPredicted]} - \text{Exp[Observed]}}{\text{Exp[Observed]}}$ |
|---|---|---|---|---|---|---|---|
| 1 | 1.41 | 3 | 0.0414 | 3 | 0.0260 | 0 | NA |
| 2 | 2.83 | 30 | −0.3432 | 15 | 0.0027 | 4 | 0.1222 |
| 3 | 5.66 | 49 | 0.1718 | 61 | −0.0332 | 53 | −0.3353 |
| 4 | 11.31 | 173 | 0.0313 | 203 | −0.2081 | 114 | 0.0757 |
| 5 | 22.63 | 295 | 0.6108 | 345 | 0.4853 | 294 | 0.3798 |
| 6 | 45.25 | 907 | 0.0405 | 1126 | −0.0251 | 984 | −0.0283 |
| 7 | 90.51 | 1699 | −0.0619 | 2382 | −0.0939 | 1953 | −0.0742 |
| 8 | 181.02 | 2471 | −0.0936 | 3528 | −0.0785 | 2819 | 0.0136 |
| 9 | 362.04 | 3010 | −0.0748 | 4374 | −0.0065 | 3723 | 0.0346 |
| 10 | 724.08 | 3013 | −0.0479 | 4646 | 0.0353 | 3805 | 0.1122 |
| 11 | 1448.15 | 2802 | −0.0552 | 4492 | 0.0219 | 3665 | 0.1182 |
| 12 | 2896.31 | 2149 | −0.0304 | 3981 | −0.0342 | 3554 | −0.0596 |
| 13 | 5792.62 | 1422 | −0.0310 | 2899 | −0.0452 | 2807 | −0.1494 |
| 14 | 11585.20 | 818 | −0.0781 | 1556 | 0.0500 | 1599 | −0.1296 |
| 15 | 23170.50 | 354 | 0.0869 | 697 | 0.1504 | 713 | −0.0429 |
| 16 | 46341.00 | 144 | 0.0299 | 319 | −0.0445 | 322 | −0.1513 |
| 17 | 92681.90 | 32 | 0.1188 | 83 | −0.1109 | 77 | 0.2686 |
| 18 | 185364.00 | 8 | −0.0685 | 14 | 0.0660 | 16 | 0.2423 |
| 19 | 370728.00 | 1 | 0.0095 | 4 | −0.0095 | 3 | −0.1535 |
| 20 | 741455.00 | 0 | NA | 0 | NA | 1 | 0.0202 |

**Figure 19. Comparison of Predicted to Observed Response for the 4-Parameter Model in 2000, 2002, 2004**

We show the results of the 3-step KS-extra sum of squares procedure, making the leap of faith that the MEPS sample approximates a sample drawn at random from the true distribution. The MEPS sample is, after all, the only one we have available to work with and insight can be gleaned for future research. It is unknown how good an approximation this is, and our analysis does not attempt to estimate or correct for the sample bias.

The results of the Kolmogorov-Smirnov goodness-of-fit test applied to the classic lognormal distribution are shown in Figures 20 and 21. A Monte Carlo was performed with 2500 repetitions to calculate the p-values.[33] The test shows the classic lognormal cannot be rejected at p=.05 in 2 out of 3 years and therefore is a plausible model.[34] The borderline nature of the results, however, is a clear indication that an improved model would be a better match to the data.

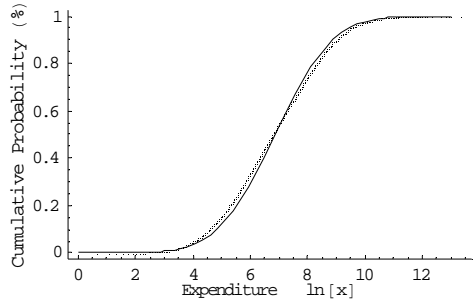

**Figure 20. Empirical Distribution Function Compared to CDF of the Classic Lognormal in 2004**

| Can we reject the hypothesis that the classic lognormal generated the observed distribution? | 2000 | 2002 | 2004 |
|---|---|---|---|
| KS Statistic | 0.019 | 0.026 | 0.034 |
| p-value (2500 repetitions) | 0.243 | 0.054 | 0.024 |
| Result | Not reject at p = .05 | Not reject at p = .05 | Reject at p = .05 |

**Figure 21. Kolmogorov-Smirnov Test Applied to Classic Lognormal**

The results of the extra sum of squares test for the 3 and 4 parameter models are shown in Figure 22. The p-value in this case is the probability that the extra sum of squares is greater than or equal to the observed amount based on pure chance, assuming the hypothesis that the additional parameter equals zero is true. A significance level of .05 means the extra sum of squares result has at most a 5% chance of being as large as observed. The test shows that the 3-parameter model is preferred to the 2-parameter classic lognormal. The situation is somewhat weaker for the 4-parameter model compared to the 3-parameter, allowing some discretion in deciding which to use. When viewed up close, healthcare practitioners have indicated that the real world exhibits the complex dynamics of the 4-parameter model.[35]

| | 2000 | 2002 | 2004 |
|---|---|---|---|
| **Comparison of 3-parameter model to classic lognormal:** | | | |
| Can we reject the hypothesis that $\delta$=0? | | | |
| Extra Sum of Squares | 0.60 | 0.11 | 1.36 |
| F-Ratio | 18.00 | 2.72 | 28.39 |
| p-Value assuming the Hypothesis is true | 0.0007 | 0.1199 | 0.0001 |
| Result | Reject at p = .05 | Not reject at p = .05 | Reject at p = .05 |
| **Comparison of 4-parameter model to 3-parameter:** | | | |
| Can we reject the hypothesis that $\epsilon$=0? | | | |
| Extra Sum of Squares | 0.01 | 0.34 | 0.19 |
| F-Ratio | 0.25 | 17.05 | 4.98 |
| p-Value assuming the Hypothesis is true | 0.6236 | 0.0010 | 0.0424 |
| Result | Not reject at p = .05 | Reject at p = .05 | Reject at p = .05 |

**Figure 22. Extra Sum of Squares Test Applied to 3 and 4-Parameter Models**

## 5. Predicted Behaviors and Implications
*Emergence of Power Law Behavior*:
A power law distribution can be used to approximate the population frequency for a fairly wide range of healthcare expenditures. The chance that extreme values can occur is high when power law behavior is observed. We will a) show how power law behavior comes about, b) make a rough estimate of the power law distribution parameters visually, and c) provide a careful estimate using Monte Carlo techniques following the approach of Clauset, Shalizi, and Newman (2007).[36] The population we will focus on is the top 1% with the largest expenditures, representing 20%-30% of total spending on healthcare. This is a highly influential group, and one that many Americans could fall into in the course of their lifetime.

Conceptually, we make a straight line approximation to the curve for $\ln[f(x)]$ at a point x=a in the region representing the top expenditures (Figure 23).

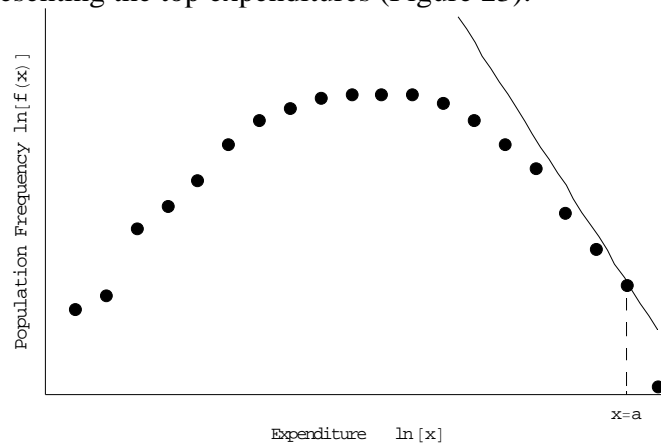

**Figure 23. Straight Line Approximation Leads to Power Law**

This means for some region $|x-a|<\zeta$
```
ln[f(x)] = -λ ln[x] + b
```
Which gives the power law
```
f(x) = cx^-λ           where c = e^b
```
The slope of the line approximates the exponent $\lambda$ of the power law. We have used the frequency function rather than the PDF, which differs by a shift, for easy comparison with the previous analysis. The power law behavior of the PDF implies the CDF in this region follows a power law as well with exponent $\gamma=-(\lambda-1)$.

The equations for the slope of the classic lognormal and the 3 and 4-parameter models using the best fit parameters for 2004 are plotted in Figure 24 for x>$10,000. This shows that a variation in x of as much as an order of magnitude changes the slope by about 1 in the region $10,000<x<$300,000.

So even though the exponent $\lambda$ of the power law is not fixed, it is approximately constant for a fairly wide range of x. Of greatest interest is the far tail, but unfortunately the PDF has so few data points at the extreme that it calls into question the reliability of the estimate of the exponent in this region of the curve. The CDF would be a much better predictor. We only have an explicit expression for the CDF, however, for the classic lognormal which we have shown is not a good fit in the far tail (Figure 3). Fortunately, there is a way around this problem.

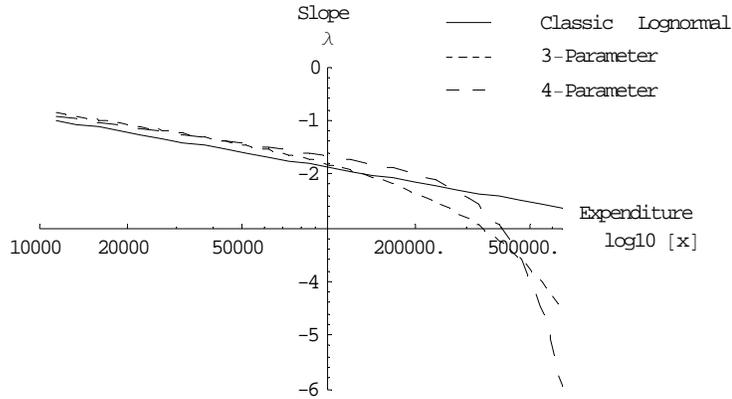

**Figure 24. Slope λ of the Curve for ln[f(x)] in 2004**

The phrase "the r[th] (r = rank, where r is a positive integer) individual, ordered from largest to smallest expenditure incurred, has expenditure x" is equivalent to saying "r individuals have expenditures equal to or greater than x". So r divided by the total number of individuals is equivalent to 1-CDF(x). A plot of $\log_{10}[x]$ against $\log_{10}[r]$ for 2004 is shown in Figure 25 where we use the base 10 logarithm instead of the natural logarithm to make it easier to interpret. This has no effect on the slope of the curve. A straight line is drawn from a linear regression of the first five data points for 2004, which roughly corresponds to the individuals that comprise the top 1% of expenditures.

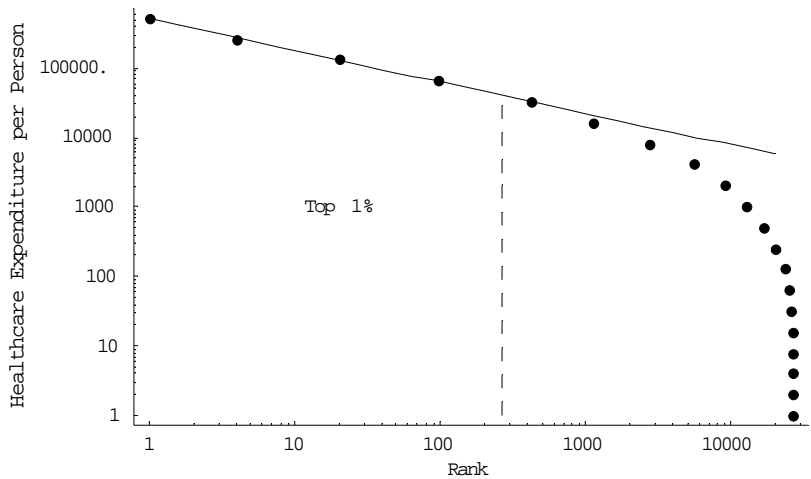

**Figure 25. Straight Line Approximation to the Curve for Healthcare Expenditure per Person vs. Rank in the Sample Population, in the Region Representing the Top 1% of Expenditures for 2004**

Making a straight line approximation to the rank plot of ln[x] against ln[r], lets write

$$\ln[x] = -\frac{1}{(\lambda - 1)} \ln[r] + \text{constant}$$

But this means
$$x \propto r^{-\frac{1}{(\lambda-1)}}$$
Inverting gives
$$r \propto x^{-(\lambda-1)}$$
We know from our previous discussion that
$$\frac{r}{N} = 1 - \text{CDF}(x)$$
The exponent for the CDF must therefore be the inverse of the slope in Figure 25
$$-(\lambda - 1)$$
The rank plot, therefore, gives an estimate of the exponent for the CDF in the far tail, $\gamma=-(\lambda-1)$, and so also for the PDF, $\lambda$. A linear regression gives the best fit exponent and the result is shown in Figure 26. The approach we have used so far has been essentially a visual one through our subjective choice of the first five data points based on a picture. There are additional problems with this estimate as well,[37] so the accuracy is low and any close match to the true value involves some luck.

The previous analysis, however, aids in understanding how to make a careful estimate of the exponent of the power law and the region where it applies, using the full data set without logarithmic binning for the top 1% of the population with the largest expenditures. This involves maximum likelihood, K-S statistics, and Monte Carlo techniques. The procedure can be summarized in three steps as follows. *First*, for each possible choice of $x_{min}$ marking the beginning of the tail, we determine the best fit estimate of the exponent $\gamma$ for the power law using the method of maximum likelihood and calculate the Kolmogorov-Smirnov goodness-of-fit statistic. We then choose the value of $x_{min}$ that gives the best fit for $\gamma$ over all values of $x_{min}$. *Second*, the uncertainty in the estimates is calculated by a bootstrap with 1000 repetitions. A synthetic data set with a similar distribution to the original data is drawn at random from the original data during each repetition. The best fit estimates of $x_{min}$ and $\gamma$ are calculated for each synthetic data set. The standard deviation of these estimates gives the estimated uncertainty in the original estimates of $x_{min}$ and $\gamma$. *Third*, a p-value is calculated to test the hypothesis that the tail is drawn from a power law distribution for $x \geq x_{min}$. This is done by generating 2500 synthetic data sets drawn from the best fit power law distribution. Each synthetic data set is fitted to a power law distribution using the methods just described above. The K-S statistic is calculated and the fraction that is larger than the K-S statistic for the original data is the p-value. The number of repetitions is high enough to provide acceptable accuracy. A small p-value such as $p \leq .05$ rejects the hypothesis that the tail is drawn from a power law distribution.

The power law parameter estimates, including the number of observations in the tail, $n_{tail}$, using the least squares visual approach and the maximum likelihood plus K-S statistics approach, are shown in Figure 26.

| Method | Parameter | 2000 | 2002 | 2004 |
|---|---|---|---|---|
| a. Least Squares | $\gamma$ | 2.502 ± 0.17 | 2.266 ± 0.056 | 2.203 ± 0.03 |
| Visual | $x_{min}$ | 32768. | 32768. | 32768. |
| Visual | $n_{tail}$ | 4 | 4 | 5 |
| b. Maximum Likelihood | $\gamma$ | 2.43 ± 0.34 | 2.47 ± 0.23 | 2.25 ± 0.16 |
| Minimum K–S Statistic | $x_{min}$ | 43967 ± 10400. | 46752 ± 8271. | 39753 ± 6260. |
| Minimum K–S Statistic | $n_{tail}$ | 110 ± 23. | 236 ± 46. | 289 ± 40. |

**Figure 26. Power Law Best Fit Parameters with Standard Errors (a) and Standard Deviations (b) in the Region Representing the Top 1% of Expenditures**

The goodness of fit test for the power law distribution is tabulated in Figure 27. The results show the power law distribution is plausible for the tail of observations where $x \geq x_{min}$

| Can we reject the hypothesis that a power law generated the observed distribution? | 2000 | 2002 | 2004 |
|---|---|---|---|
| p-value (2500 repetitions) | 0.13 | 0.67 | 0.56 |
| Result | Not reject at p = .05 | Not reject at p = .05 | Not reject at p = .05 |

**Figure 27. Goodness-of-Fit Test Applied to the Power Law Distribution in the Region Representing the Top 1% of Expenditures**

A comparison of the power law distribution to the classic lognormal using the likelihood ratio test for non-nested models shows that while the power law is preferred, the test is not sensitive enough to discriminate between the two distributions (Figure 28). The two distributions are almost equal over the range of x examined and a larger sample size is needed to tell them apart.[38]

| Comparison of Power Law Model with Classic Lognormal | 2000 | 2002 | 2004 |
|---|---|---|---|
| Log Likelihood Ratio | -0.54 | -4.9 | -4.573 |
| Is the power law model preferred to the classic lognormal model? | Yes | Yes | Yes |
| p–Value | 0.832 | 0.193 | 0.154 |
| Is the likelihood ratio test adequate to discriminate between the models? | No | No | No |

**Figure 28. Likelihood Ratio Test Comparing the Power Law Distribution to the Classic Lognormal in the Region Representing the Top 1% of Expenditures**

These results explain why researchers have previously proposed using a Pareto (i.e., power law or Zipf) distribution to fit the upper tail of 1-CDF.[39]

*Highest Cost Medical Conditions*
The most expensive medical conditions should be concentrated in the tail of the distribution where power law behavior is exhibited. So it is reasonable to expect that the cost of treatment of high cost medical conditions should follow a power law. Indeed, a linear regression of the mean annual treatment costs per person for the most expensive medical conditions in America yields a power law (Zipf) relation with an $R^2 = .96$ [40] The results are shown in Figure 29 with the top five conditions being respiratory malignancies, ischemic heart disease, congestive heart failure, cerebrovascular disease, and motor vehicle accidents.

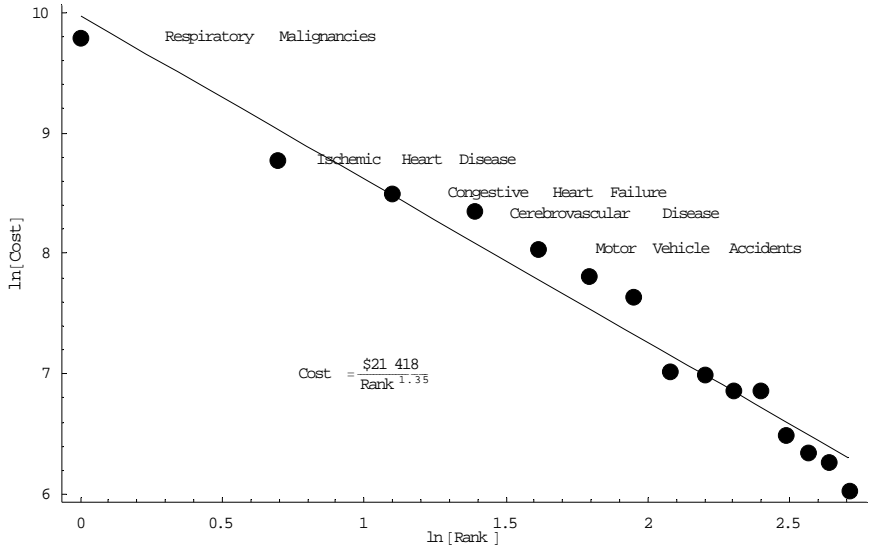
**Figure 29. Most Expensive Medical Conditions in the US**

*Connection with Mathematical Finance*:
A fundamental connection exists between healthcare expenditures and mathematical finance. This can be seen by exploring the time-dependent behavior of the model. If we return to the original equation and make the parameters a function of time we have

$$\ln[x(t)] - \ln[x_0] = \delta(t)\, x + \epsilon(t)\, x^2 + N[\mu(t), \sigma^2(t)]$$

So, as others have described, we have a kind of "traveling normal distribution" [41] Examination of the data from 1996 to 2004 reveals the actual form of this time dependency. For simplicity we will limit the discussion to the 3-parameter model, and since there are 9 years of annual data rather than decades of data at a monthly or daily time-scale, we will focus on the long-term low resolution behavior seen.

The mean, $\mu(t)$, is well-approximated by a straight line, with cyclic behavior about the trend. A linear regression gives an $R^2 = .97$ for the best fit function

$$\mu(t) = 0.08131\, t - 155.959$$

where t is a year in the range [1996, 2004]. This corresponds to a continuously compounded annual growth rate of 8.1% for expenditures.[42] A plot of the data and best fit function for $\mu(t)$ is shown in Figure 30

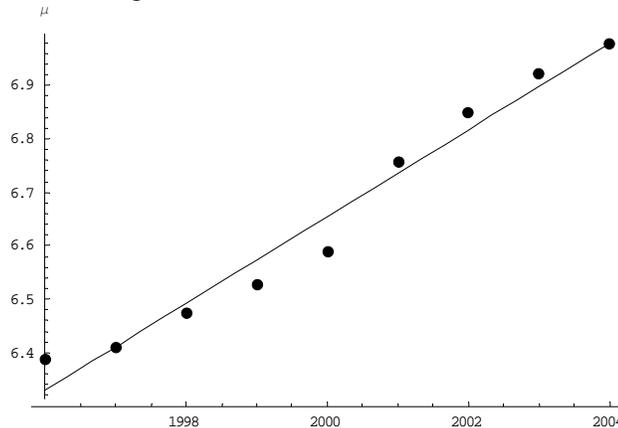
**Figure 30. Mean of Healthcare Expenditures $\mu(t)$**

The cyclic behavior of the mean can be seen by plotting the deviation from trend. The mean spending on healthcare appears to be periodic based on slightly more than one cycle of observations. A nonlinear regression gives this best fit function

    µ (t) = -155.966 + 0.08131 t - 0.046 Sin[1.0002 t]

A plot of the deviations is shown in Figure 31.

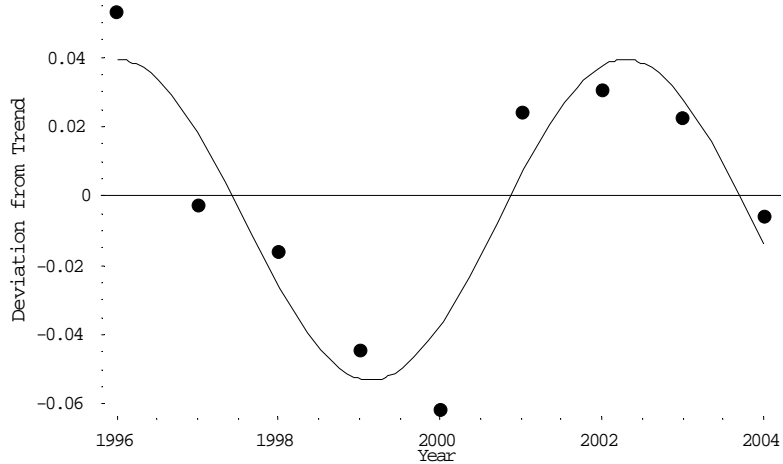

**Figure 31. Deviation of the Mean *µ*(t) from Trend**

The infusion term, *δ*(t), has been mostly negative in the past reflecting economies of scale in treating cases of serious illness, but has been positive occasionally as shown in Figure 32. In contrast to Figure 30 where the points are actual observations, here they are fitted values. There may be some damped periodicity in *δ*(t), but the standard errors in the estimates of *δ*(t) are large and the time series is short. So approximating *δ*(t) by its average value is reasonable.

    δ (t) = -1.03 × 10⁻⁶

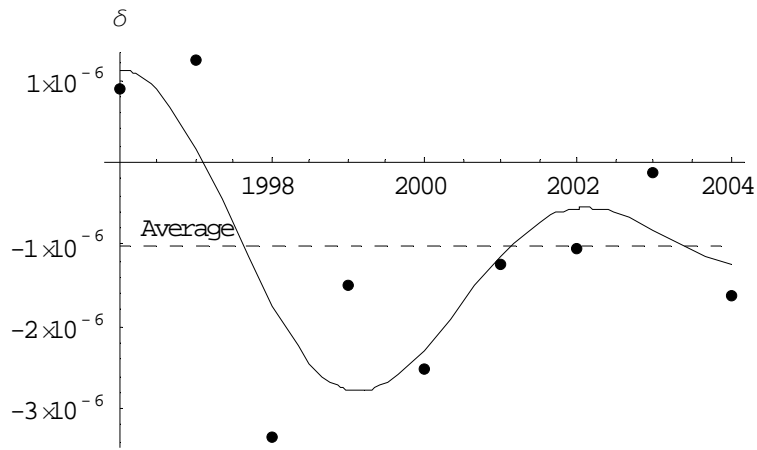

**Figure 32. Coefficient of Infusion *δ*(t)**

The variance in consumer spending on healthcare, $\sigma^2(t)$, also appears to be periodic though it is not as clear due primarily to the data point in 2000. A nonlinear regression gives this best fit function

$$\sigma^2(t) = 2.656 - 0.025 \sin[0.9993\, t]$$

A plot is shown in Figure 33. Periodicity is apparent in the unweighted variance data.

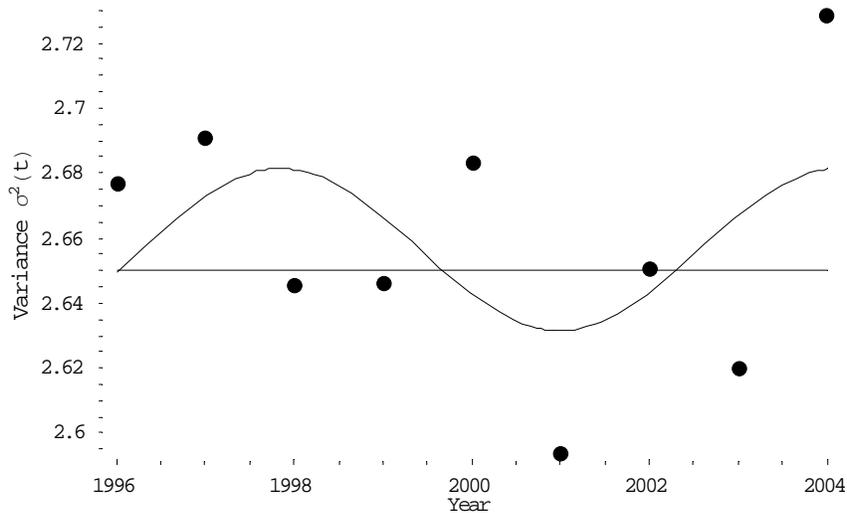

**Figure 33. Variance of Healthcare Expenditures $\sigma^2$(t)**

The period of the mean and variance is nearly identical to the business cycle, lasting 6.3 (approximately $2\pi$) years and differing by a fraction of a percent. The underwriting cycle that has been observed in the healthcare insurance industry over the decades also has a 6-year period.[43] A plot of the deviation of the mean, variance, and GDP from trend [44] with the last two time-shifted, is shown in Figure 34.

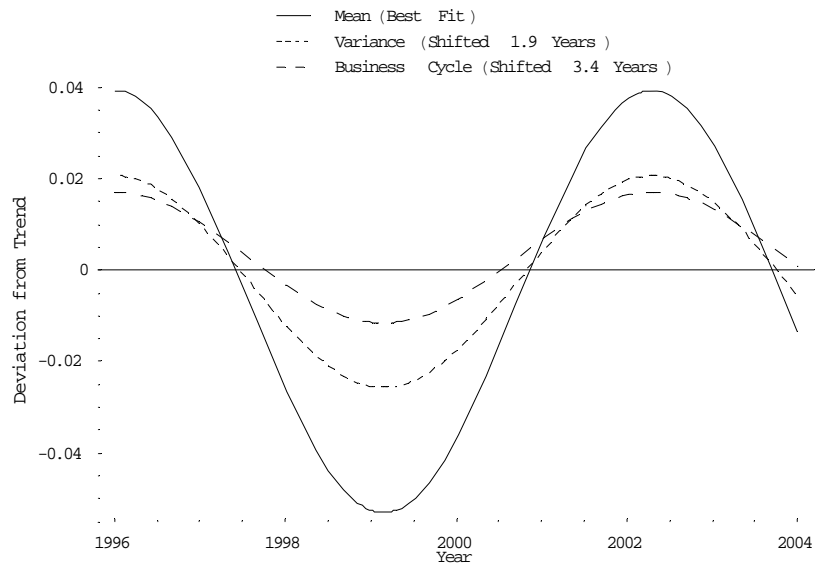

**Figure 34. Comparison of Mean and Variance to the Business Cycle**

When the variance is detrended, the best fit function can be separated into a periodic deviation about a quadratic trendline. This refines the expression for $\sigma^2(t)$, and gives:
`σ² (t) = 0.000407838 t² – 1.63216 t + 1635.633 – 0.023 Sin[0.9993 t]`
A plot of the time average of the variance, $<\sigma^2(t)>$, is shown in Figure 35.

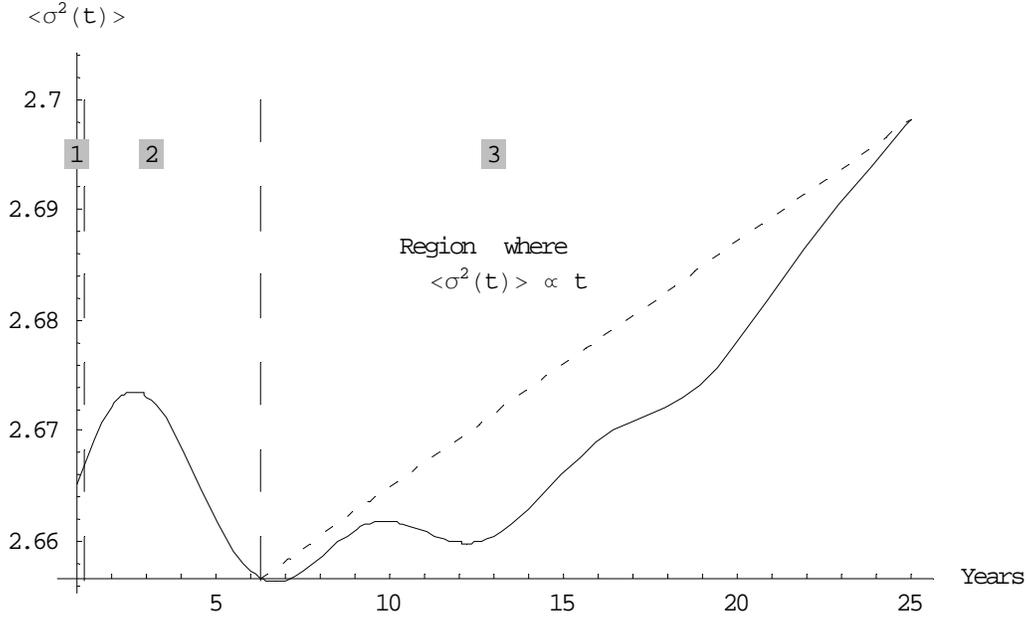

**Figure 35. Time Average of Variance (Year 1=1996)**

Assuming the period 1996-2004 is representative of the future, we can separate the healthcare expenditure process into 3 regions based on the time average: 1) T ≤ 1 year, 2) 1 year < T ≤ 6.3 years, and 3) T > 6.3 years, where T = t-t$_o$ is the length of a time interval and 6.3 years is the period of the mean and variance. To simplify expressions, we will set the clock so that Year 1=1996, and therefore T=t in the following.

For T ≤ 1 year, we can consider $\mu(t)$, $\sigma^2(t)$ constant and our previous results hold.
`ln[x] – ln[x₀] = δ x + N[μ, σ²]`
For 1 year < T ≤ 6.3 years, $\mu(t)$ is linear and $\sigma^2(t)$ is sinusoidal. We need to modify our original time-independent results. Substituting the best fits for $\mu(t)$ and $\sigma^2(t)$ into the original equation for ln[x] gives

```
ln[x (t)] – ln[x₀] = δ (t) x + N[μ (t), σ² (t)]
 = –1.03 × 10⁻⁶ x + .08131 t + 6.329
   N[0, .000408 t² – .00408 t + 2.664 – .023 Sin[0.9993 t + .8841 π]]
```

For T > 6.3 years, the long-term trends are predominant and so the time average <ln[x(t)]> is of interest. The additive property of the normal distribution allows us to take the time averages of the parameters and so we have
`< ln[x (t)] > –ln[x₀] = < δ (t) > x + N[ < μ (t) >, < σ² (t) >]`
Where
`< δ (t) > = δ = –1.03 × 10⁻⁶`

$$<\mu(t)> = \frac{1}{t-t_o} \int_{t_o}^{t} \mu(t)\, dt = 6.329 + 0.0407\, t$$

$$<\sigma^2(t)> = \frac{1}{t-t_o} \int_{t_o}^{t} \sigma^2(t)\, dt =$$

$$= 2.664 - 0.00204\, t + 0.000136\, t^2 +$$

$$\frac{0.023}{t}\left(\cos[0.9993\, t + 0.884111\, \pi] + 0.94157\right)$$

Over this longer time interval the process healthcare expenditures follow becomes similar to a widely used model for managing financial assets. This is because the time average of the variance can be approximated by a straight line after one cycle of 6.3 years so that $\sigma^2(t) \propto t$.

$$<\sigma^2(t)> = 2.671 + 0.002\, t$$

The equation for $<\ln[x(t)]>$ is then,

$$<\ln[x(t)]> - \ln[x_0] = \delta x + N\left[<\dot\mu(t)> t,\ <\dot\sigma^2(t)> t\right] + N[\mu_o,\ \sigma_o^2]$$

$$\text{where } <\dot\mu(t)> = \frac{d<\mu(t)>}{dt},\quad <\dot\sigma^2(t)> = \frac{d<\sigma^2(t)>}{dt}$$

and $\mu_o = <\mu(0)>,\ \sigma_o^2 = <\sigma^2(0)>$

We can change this to an annual equation by taking the time derivative of t multiplied by the time average[45] and letting

$$\mu_* = 2<\dot\mu(t)>,\quad \sigma_*^2 = 2<\dot\sigma^2(t)>$$

which gives

$$\ln[x(t)] - \ln[x_0] = \delta x + N[\mu_* t,\ \sigma_*^2 t] + N[\mu_o,\ \sigma_o^2]$$

$$x(t) = x_0\, e^{\delta x + N[\mu_* t,\ \sigma_*^2 t] + N[\mu_o,\ \sigma_o^2]}$$

with x(t) representing the healthcare expenditure of an individual in year t; $\delta x$ a discount (or premium) as a result of economies (or diseconomies) of scale; $\mu_*$ the continuously compounded annual rate of growth of expenditures (about 8.1%); $\sigma_*^2$ the variance of the continuously compounded annual rate of growth of expenditures per year (about .4% per year or 6.3% per $\sqrt{\text{year}}$ for the standard deviation $\sigma_*$)[46]; and $N[\mu_o, \sigma_o^2]$ the minimum fluctuation in healthcare expenditures for the individuals in the US population with non-zero expenditures at $t_o$

This has the same form as Brownian motion with drift, widely used to model the price of stocks[47]

$$\ln[S(t)] - \ln[S_0] = N[\mu t,\ \sigma^2 t]$$

$$S(t) = S_0\, e^{N[\mu t,\ \sigma^2 t]}$$

Here S(t) is the price of a stock at time t and, in this context, $\mu$ is the rate of return and $\sigma$ is the volatility, i.e., the average and standard deviation, respectively, of the logarithm of the growth in the stock price for a given time period. The difference between the equations is simply the infusion term $\delta x$ and a baseline fluctuation $N[\mu_o, \sigma_o^2]$. Were these terms to be included in the model for stocks, it would simply mean there is a relation between the growth in price and the price level of a stock, and that there is a minimum fluctuation at $t_o$. The model of stock prices, however, assumes that growth in price is not dependent on current price, and is meant to describe the behavior of a single stock which has a definite price at $t_o$, so neither of these terms appears. Were a single individual to be considered in the healthcare expenditure model and $\delta=0$ then the equations would be essentially the same.

Doctors, in this regard, can therefore be viewed as managers of human capital comprised of a portfolio of patients, in analogy with investment managers who manage financial capital comprised of a portfolio of securities. Similarly, patients are owners of a single "stock of health", which is managed by themselves or fiduciaries.

*Sample Path*
A sample path for the total healthcare expenditure in the US can be plotted. Given that we are dealing with macroscopic quantities in the form of a significant amount of expenditure, large population, and an entire year, plotting a sample path is an unconventional application of a technique usually reserved for microscopic phenomenon. A helpful way to think about it is as a population bombarded by $>10^8$ cases of illness per tick of the clock.

The task of plotting a sample path for the total healthcare expenditure in the US involves 4 steps.
- Take a time period T and divide it into k intervals, $T=k\Delta t$ where $\Delta t = 1$ year
- For each time interval k draw $n_{pop}$ samples from the distribution of healthcare expenditures for that year. The number $n_{pop}$ should reflect the growth in population from year to year
- Add up all the expenditures to get the total annual healthcare expenditures in the US for each year
- Plot the points {k, total annual healthcare expenditure}

The second step is where the work is. There is no random number generator available that can easily produce the results we need. However, we do have an efficient way to generate random numbers for a lognormal distribution and a power law distribution. By splicing these together, say 99% of the population is drawn from a lognormal and 1% from a power law, we have a way to approximate the true distribution.

A similar procedure could be applied to a single individual to model healthcare expenditures over the course of a lifetime, where a sample is drawn with 99% probability from a lognormal and 1% from a power law. This raises, however, the issue of correlations in the data. The healthcare expenditure model assumes that individuals start each year with a clean bill of health. Chronic illness, risky behaviors, and disadvantages, however, lead to persistence in healthcare expenditures. It would be useful to know the correlations in the data to sort this out for individuals. The long-hoped for electronic medical record, if designed properly, might provide this someday. In the meantime, the model can provide a sample path for the abstract "average" American.

*Metrics and Benchmarks*
The framework of an annual performance measure for healthcare cost control is $(\mu+\delta x+\epsilon x^2)\,\sigma$ which relates each of the parameters in the model. For simplicity we will focus on the case of 3 parameters with $\epsilon=0$.

We want the measure to be easy to interpret, as well as independent of x so its value incorporates the cumulative effect of $\delta$ across the range of expenditures. This can be done by 1) calibrating the measure to reflect the tradeoff between $\mu$ and $\delta$, and 2) changing the scale to represent the discount/premium on total healthcare expenditures due to $\delta$. The result, as we will see, is the arithmetic measure $\alpha\beta(1+\delta x_{tradeoff})^2$, where $\alpha$ is the average annual healthcare expenditure, $\beta$ the standard deviation (i.e., risk), and $\delta x_{tradeoff}$ the percent discount/premium on healthcare expenditures due to $\delta$.

The requirement that the measure be independent of x can be stated mathematically as
$$(\mu + \delta x)\sigma = \text{constant}$$
which has the solution
$$x = -\frac{\partial \mu}{\partial \delta}$$
This quantity can be determined by calculating the change in the total cost of healthcare for the entire population when $\mu$ changes by $\partial \mu$ and $\delta$ by $\partial \delta$. The ratio calibrates the tradeoff between $\mu$ and $\delta$.

$$\frac{\partial \mu}{\partial \delta} = \frac{\frac{\partial}{\partial \delta}\int_1^{x_{max}} x\, f_3(\mu, \delta, \sigma, x)\, dx}{\frac{\partial}{\partial \mu}\int_1^{x_{max}} x\, f_3(\mu, \delta, \sigma, x)\, dx}$$

This can be solved numerically for a given point in parameter space $\{\mu, \delta\}$. Figure 36 shows the ratio, $x_{tradeoff}$, evaluated at the best fit parameters

$$x_{tradeoff} = -\left(\frac{\partial \mu}{\partial \delta}\right)_{\text{best fit}}$$

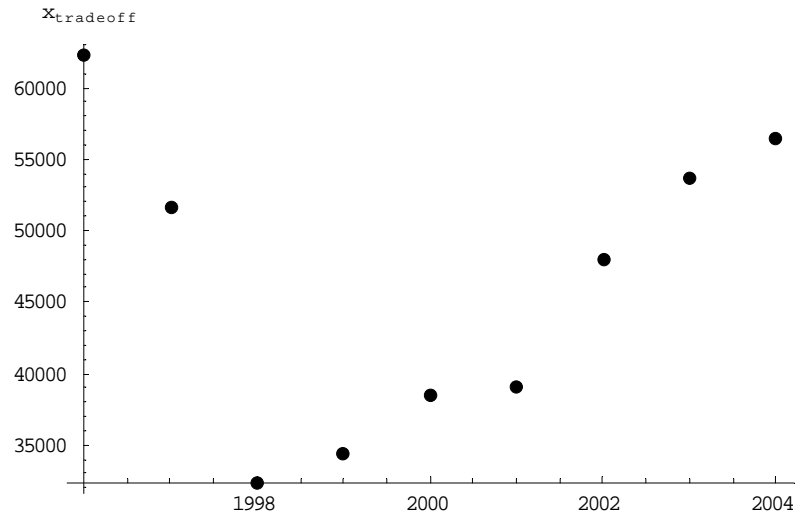

**Figure 36.** $x_{tradeoff}$

The sensitivity of $x_{tradeoff}$ to a change in $\mu$ or $\delta$, which might be attempted by a cost control effort, is the subject of Figure 37. From the plots shown and the relation

$$\left(\frac{\partial \mu}{\partial \delta}\right)_{\text{best fit}} = \left(\frac{\frac{\partial \text{Total Healthcare Cost}}{\partial \delta}}{\frac{\partial \text{Total Healthcare Cost}}{\partial \mu}}\right)_{\text{best fit}}$$

we can deduce that a 10% change in $x_{tradeoff}$ can be caused by a 1.5% change in $\mu$ or a 20% change in $\delta$.

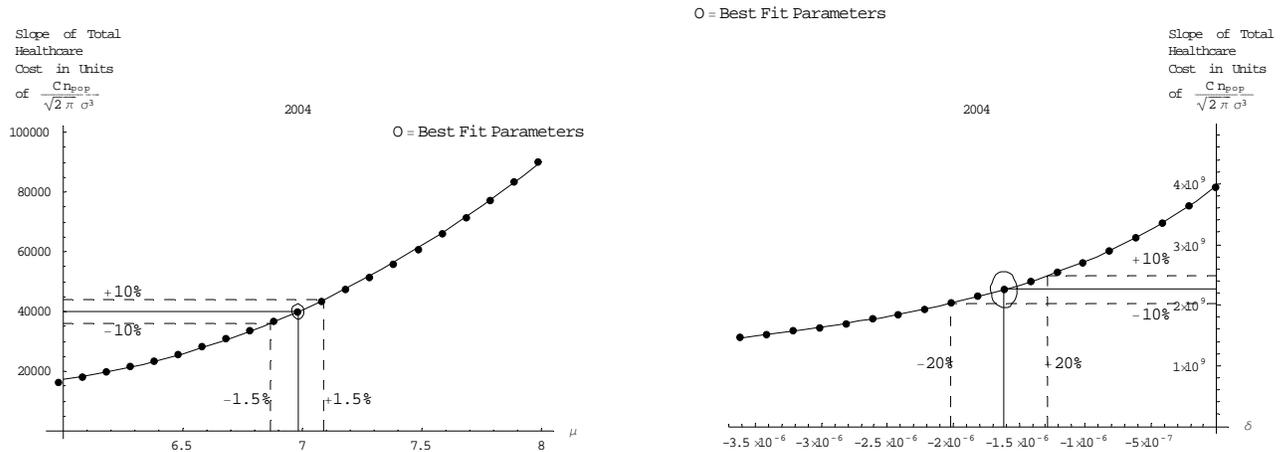

**Figure 37. Sensitivity of $x_{tradeoff}$**

The annual performance measure for healthcare cost control can now be calibrated to reflect the tradeoff between $\mu$ and $\delta$

$$(\mu + \delta\, x_{tradeoff})\, \sigma = \left(\mu - \delta\, \frac{\partial \mu}{\partial \delta}_{best\ fit}\right) \sigma$$

Putting $\Delta\delta = \delta - 0$ we can write

$$\left(\mu - \Delta\delta\, \frac{\partial \mu}{\partial \delta}_{best\ fit}\right) \sigma = \mu \left(1 - \frac{\Delta\mu}{\mu}\right) \sigma$$

This should look familiar and is a result we could have guessed; it is the average of the logarithm of healthcare expenditures multiplied by one minus the logarithmic discount/premium, leaving a *net amount*, multiplied by the standard deviation of the logarithm of x. In other words, it is the expression found earlier for estimating the effective value of the logarithmic mean assuming the classic lognormal model, $\mu(\delta=0)$, times the logarithmic standard deviation. Plots comparing the actual to the estimates for the logarithmic mean, standard deviation, and discount/premium are shown in Figure 38 a, b, c.

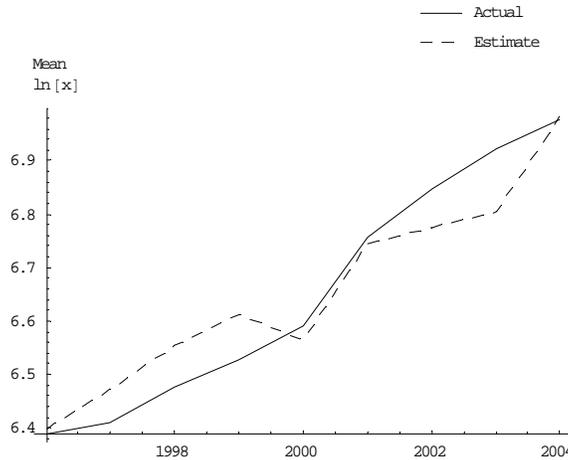

**Figure 38 a. Logarithmic Mean**

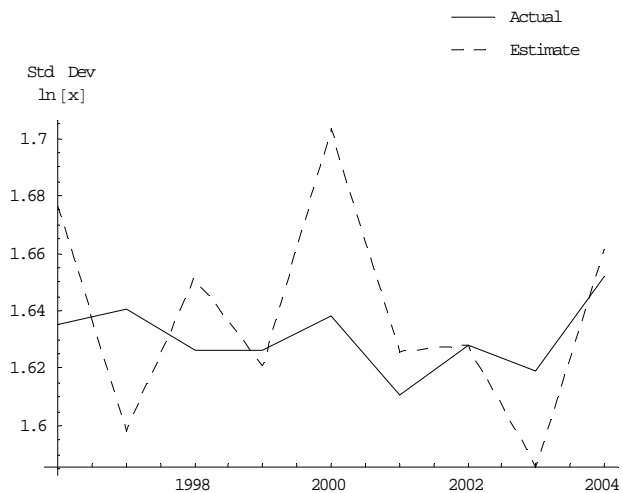

**Figure 38 b. Logarithmic Standard Deviation**

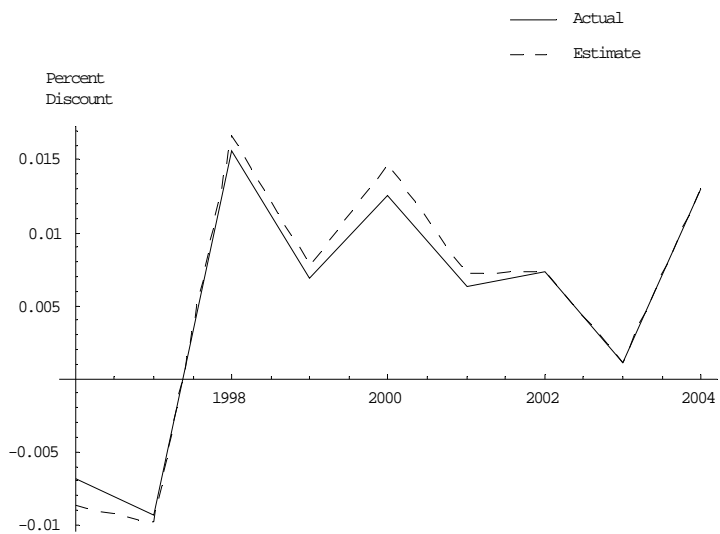

**Figure 38 c. Logarithmic Discount/Premium**

The logarithmic discount/premium is not an intuitive quantity. The arithmetic discount/premium, however, is easily understood as the percent change in total healthcare expenditure due to $\delta$.

$$\frac{\Delta \bar{x}}{\bar{x}} \approx \frac{\left(e^{\mu+\frac{\sigma^2}{2}} - e^{\mu[\delta=0]+\frac{\sigma^2}{2}}\right)\left(1 - \Delta\delta \left(\frac{\partial \mu}{\partial \delta}\right)_{\delta=\delta}\right)}{e^{\mu+\frac{\sigma^2}{2}}\left(1 - \Delta\delta \left(\frac{\partial \mu}{\partial \delta}\right)_{\delta=\delta}\right)} = \frac{n_{pop}\left(e^{\mu+\frac{\sigma^2}{2}} - e^{\mu[\delta=0]+\frac{\sigma^2}{2}}\right)}{n_{pop}\, e^{\mu+\frac{\sigma^2}{2}}}$$

$$= \frac{X(\mu) - X(\mu[\delta=0])}{X(\mu)} = \frac{\Delta X}{X}$$

where

$$X(\mu) = \int_1^{x_{max}} x\, f_2(\mu, \sigma, x)\, dx$$

It is preferable, therefore, to change the measure to an arithmetic scale. This can be done using the previously derived formulas for estimating the arithmetic mean and standard deviation. The transformation of the logarithmic measure to an arithmetic scale gives the desired result

$$\alpha \left(1 - \Delta\delta \left(\frac{\partial \mu}{\partial \delta}\right)_{\delta=\delta}\right)^2 \beta$$

$$\boxed{= \alpha \, (1 + \delta \, x_{tradeoff})^2 \, \beta}$$

where

$$\alpha = e^{\mu + \frac{\sigma^2}{2}}, \quad \beta = e^{\mu + \frac{\sigma^2}{2}} \sqrt{(e^{\sigma^2} - 1)}$$

Since the arithmetic discount/premium has typically been less than 10% (see Figure 38) notice that

$$\frac{\Delta \bar{x}}{\bar{x}} = 1 - e^{-\Delta\mu} = 1 - e^{-\delta \left(\frac{\partial \mu}{\partial \delta}\right)_{\delta=\delta}} \approx -\delta \left(\frac{\partial \mu}{\partial \delta}\right)_{\delta=\delta} = \delta \, x_{tradeoff}$$

So $\delta x_{tradeoff}$ is the arithmetic discount/premium. The series expansion is a good approximation, with less than 5% error typically.[48]

The arithmetic measure $\alpha(1+\delta x_{tradeoff})^2 \beta$ is now independent of x and easy to understand: it is *the net average annual cost of healthcare times the effective risk*. Plots comparing the actual to the estimates for the net average annual cost of healthcare (arithmetic mean), effective risk (arithmetic standard deviation), and the percent discount/premium on healthcare expenditure (arithmetic discount/premium) are shown in Figure 39 a, b, c.

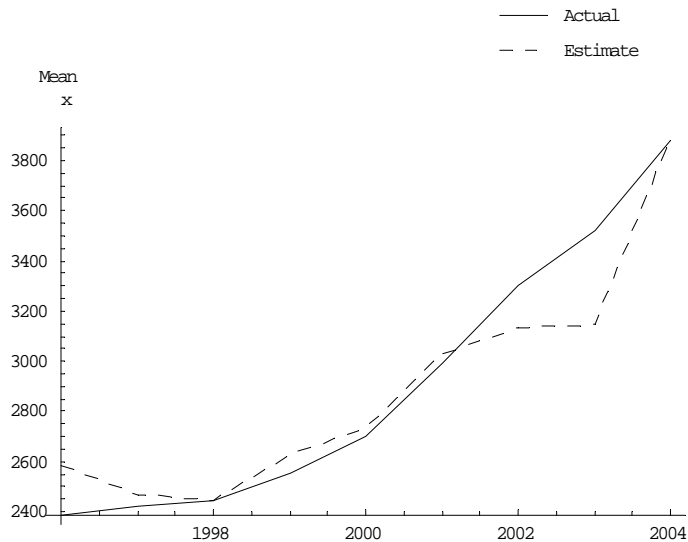

**Figure 39 a. Arithmetic Mean**

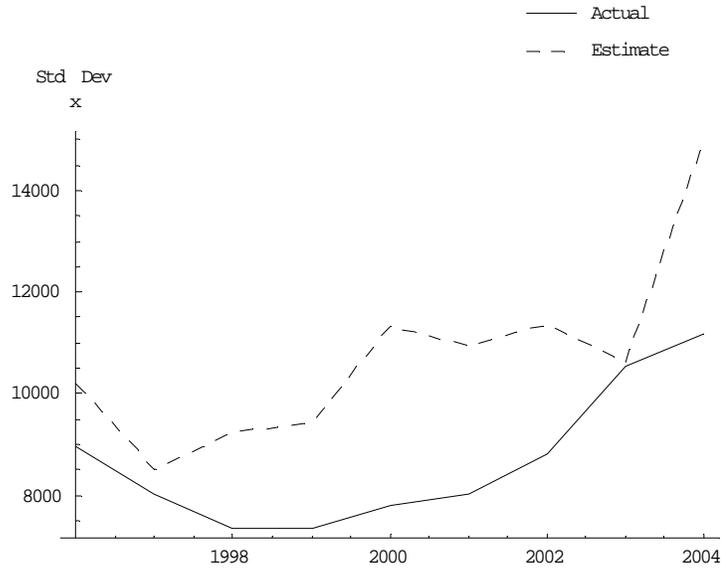

**Figure 39 b. Arithmetic Standard Deviation**

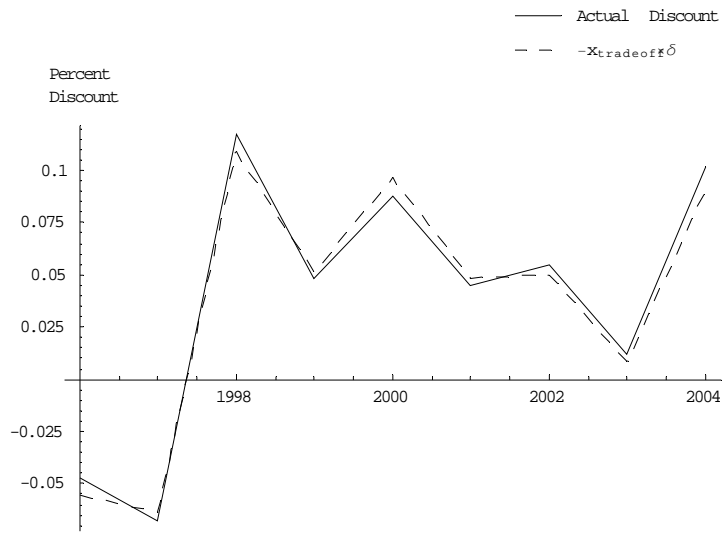

**Figure 39 c. Arithmetic Discount/Premium**

We can now return to the interpretation of the metric $(\mu+\delta x_{tradeoff})\sigma$. The expression for the metric is the product of the deterministic and probabilistic terms that drive healthcare expenditures and therefore is a measure of the cost and control of illness.

When $\mu$ and $\delta$ are both zero the entire process is probabilistic, while when $\sigma$ is zero the entire process is deterministic. When all three are zero, there is no longer any process at all. In this regard, the expression is a measure of the degree to which medicine has conquered illness, and a decrease in $\mu$, $\delta$, and/or $\sigma$, which leads to a proportional decrease in the measure, represents progress towards that goal.

The multiplication sign scales the deterministic term by the probabilistic term so that the metric moves in proportion with them and has the same direction.[49] The plus sign reflects that a change in $\mu$ must be offset by an opposite change in $\delta$ to keep the metric constant or decrease it.

When we group the triplet of parameters that comprise the metric into the pairs $\mu\sigma$, $-\delta/\mu$, and $-\delta/\sigma$ the separate tradeoffs between $\mu$, $\sigma$, and $\delta$ are highlighted. The first pair is the information factor, $\mu\sigma$,[50] which is the product of the mean and the standard deviation. It is therefore the average of the logarithm of the annual cost of healthcare, weighted by its risk. It is the major factor in overall cost control. Directionally, smaller is better.

The second pair, $-\delta/\mu$, represents the discount on the average of the logarithm of the annual cost of healthcare due to economies of scale, $\delta$. It is the major factor in cost-control for the seriously ill. The actual amount of the discount is calculated by comparing the total healthcare cost of the 3-parameter model to the classic lognormal, keeping the mean and the variance fixed. This gives Figure 40 which shows the average discount during the 5-year period 2000 to 2004 is about 6%. The correlation of $-\delta/\mu$ and the logarithmic discount is .96. Since $\mu$ is logarithmic and $\delta$ is arithmetic, the ratio $\delta/\mu$ is an index that correlates with the amount of the discount rather than a physical quantity. Directionally, bigger is better and it would be reasonable to multiply it by say $10^6$ so the numbers are easier to handle.

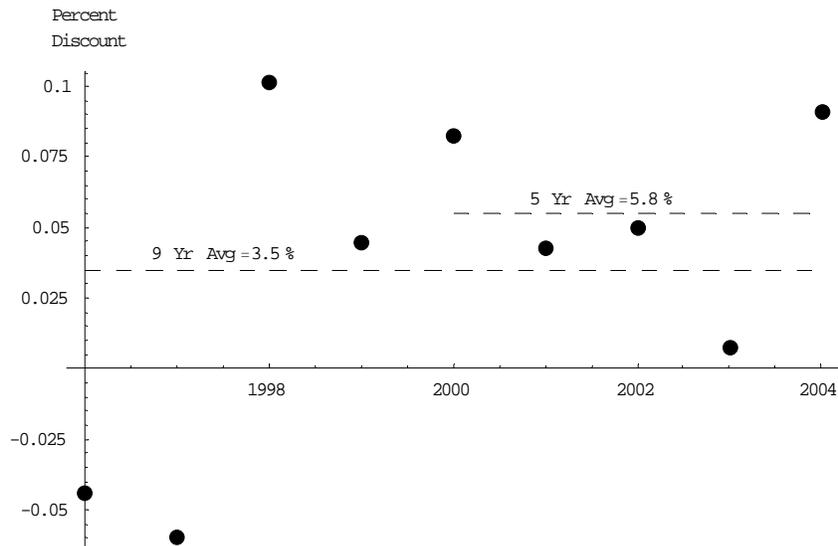

**Figure 40. Discount Due to Economies of Scale**

The third pair, $-\delta/\sigma$, is the ratio of the coefficients of infusion and diffusion, and measures the discount per unit of risk. Directionally, bigger is better and it would also be reasonable to multiply it by say $10^6$ so the numbers are easier to handle.

The parameters $\mu_*$ and $\sigma_*$ can be analyzed in a similar way to assess the annual rate of growth in cost, its volatility, and long term trends.

The arithmetic measure $\alpha(1+\delta x_{tradeoff})^2\beta$ differs from the logarithmic in a crucial aspect: changes in $\mu$ and $\sigma$ have an exponential effect. This is even more pronounced since these parameters are multiplied by a factor of two in the case of $\mu$ while $\sigma$ is squared.

$$\alpha\,(1 + \delta\, x_{tradeoff})^2\,\beta = e^{2\mu+\sigma^2}\sqrt{e^{\sigma^2}-1}\left(1 - \Delta\delta\left(\frac{\partial \mu}{\partial \delta}\right)_{\delta=\delta}\right)^2$$

Increases in $\mu$ and $\sigma$ therefore result in a sharp increase in the amount and uncertainty of healthcare expenditures. On the other hand, decreases are just as rapid.

A plot of the logarithm of the actual total healthcare expenditure compared to the estimated values of the logarithmic mean, the information factor, and the logarithmic metric is shown in Figure 41. While total costs rose from 1996-2004, the logarithmic metric and the information factor reveal that cost control improved some years (this is corroborated by substituting the actual for the estimated values of the parameters). Since the figures are not adjusted for inflation or population increases, the cost control progress during these years is even larger than shown.

A plot of the actual logarithmic discount (i.e., the ratio of the total healthcare cost of the 3-parameter model to the classic lognormal divided by $\mu$) compared to the estimates of the discount alone (i.e., the coefficient of infusion), the discount per unit of risk (i.e., the ratio of the coefficients of infusion and diffusion), the discount on the average of the logarithm of the annual cost of healthcare (i.e., the ratio of the coefficient of infusion and the logarithmic mean), and the negative logarithmic metric is shown in Figure 42.

A parallel set of arithmetic comparisons are shown in Figures 43 and 44.

The correlations of total healthcare expenditure, the actual discount, the metric, each of the parameter pairs, and the individual parameters are shown in Figures 45 and 46. Correlations of special interest are highlighted for selected measures of cost control: the metric, information factor, mean, and the infusion. They differ in degree of comprehensiveness and complexity.

The logarithmic metric, $(\mu+\delta x_{tradeoff})\sigma$, is significantly correlated with the parameter pairs, individual parameters, logarithm of total healthcare expenditure, and the logarithmic discount. The logarithmic metric is comprehensive but also complex. The information ratio, $\mu\sigma$, is significantly correlated with its constituent parameters, logarithm of total healthcare expenditure, and the logarithmic discount. It is less comprehensive but simple and easy to interpret. The logarithmic mean, $\mu$, is significantly correlated with the logarithm of total healthcare expenditure, but not with $\sigma$ or $\delta$; while the infusion, $\delta$, is significantly correlated with the actual logarithmic discount, but not with $\mu$ or $\sigma$. They are the least comprehensive but simplest.

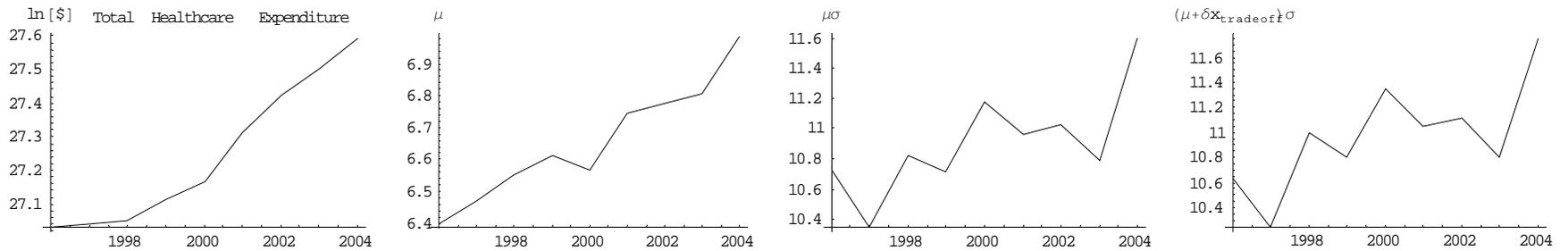

**Figure 41. Comparison of the Actual Logarithm of Total Healthcare Expenditures to Estimates of the Logarithmic Mean, Information Factor, and Positive Metric**

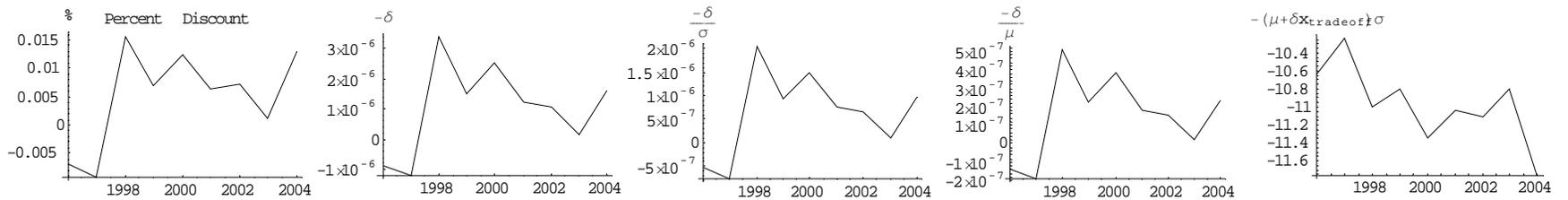

**Figure 42. Comparison of the Actual Logarithmic Discount to Estimates of the Coefficient of Infusion, Discount per unit of Risk, Discount on the Average of the Logarithm of the Annual Cost of Healthcare, and the Negative Metric**

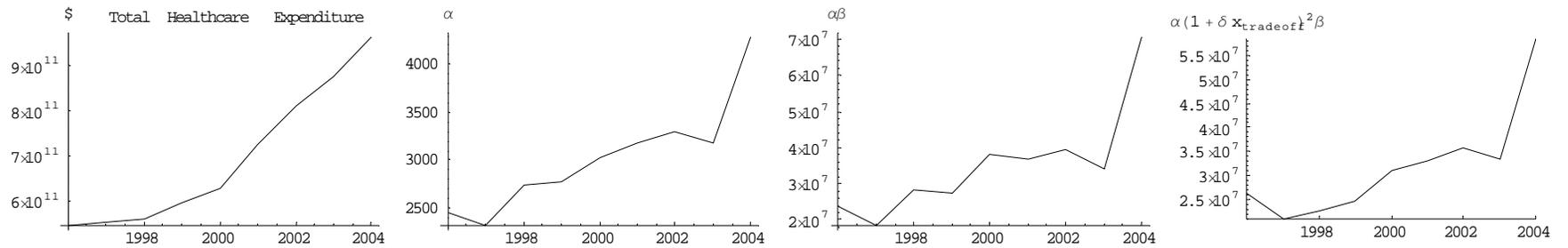

**Figure 43. Comparison of Actual Total Healthcare Expenditures to Estimates of the Arithmetic Mean, Information Factor, and Positive Metric**

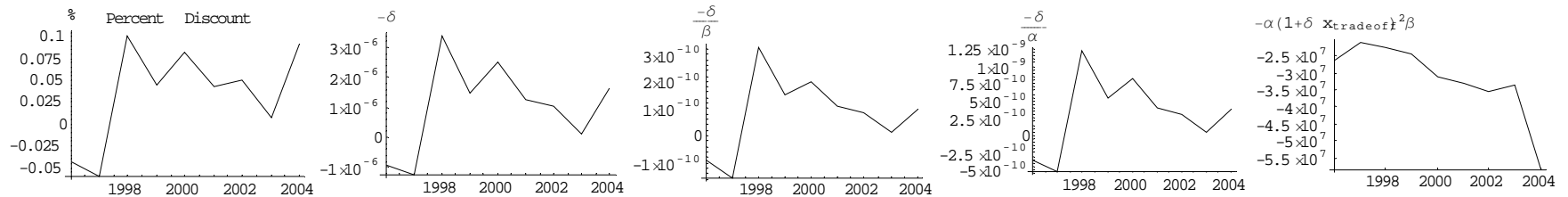

**Figure 44. Comparison of the Actual Discount to Estimates of the Coefficient of Infusion, Discount per unit of Risk, Discount on the Average Annual Cost of Healthcare, and the Negative Metric**

| Correlation | $(\mu+\delta x_{tradeoff})\sigma$ | $\mu\sigma$ | $-\frac{\delta}{\mu}$ | $-\frac{\delta}{\sigma}$ | $\mu$ | $\sigma$ | $\delta$ | $\ln \sum_{i=1}^{n_{pop}} x_i$ | $\ln$ Discount |
|---|---|---|---|---|---|---|---|---|---|
| $(\mu+\delta x_{tradeoff})\sigma$ | 1 | 0.9660 | 0.2932 | 0.3081 | 0.7038 | 0.5003 | −0.3095 | 0.7203 | 0.5087 |
| $\mu\sigma$ | | 1 | 0.5271 | 0.5403 | 0.7183 | 0.5300 | −0.5416 | 0.6833 | 0.7130 |
| $-\frac{\delta}{\mu}$ | | | 1 | 0.9995 | 0.2572 | 0.4269 | −0.9998 | 0.0825 | 0.9624 |
| $-\frac{\delta}{\sigma}$ | | | | 1 | 0.2767 | 0.4216 | −0.9998 | 0.0997 | 0.9677 |
| $\mu$ | | | | | 1 | −0.2091 | −0.2712 | 0.9591 | 0.4894 |
| $\sigma$ | | | | | | 1 | −0.4301 | −0.2046 | 0.4048 |
| $\delta$ | | | | | | | 1 | −0.0956 | −0.9670 |
| $\ln \sum_{i=1}^{n_{pop}} x_i$ | | | | | | | | 1 | 0.3279 |
| $\ln$ Discount | | | | | | | | | 1 |

Figure 45. Logarithmic Metric and Parameter Correlations

| Correlation | $\alpha(1+\delta x_{tradeoff})^2\beta$ | $\alpha\beta$ | $-\frac{\delta}{\alpha}$ | $-\frac{\delta}{\beta}$ | $\alpha$ | $\beta$ | $\delta$ | $\sum_{i=1}^{n_{pop}} x_i$ | Discount |
|---|---|---|---|---|---|---|---|---|---|
| $\alpha(1+\delta x_{tradeoff})^2\beta$ | 1 | 0.9765 | 0.0769 | 0.0812 | 0.9641 | 0.9404 | −0.1748 | 0.8762 | 0.4375 |
| $\alpha\beta$ | | 1 | 0.2761 | 0.2766 | 0.9799 | 0.9835 | −0.3735 | 0.8228 | 0.6109 |
| $-\frac{\delta}{\alpha}$ | | | 1 | 0.9980 | 0.2865 | 0.3720 | −0.9929 | 0.0158 | 0.9272 |
| $-\frac{\delta}{\beta}$ | | | | 1 | 0.2930 | 0.3695 | −0.9884 | 0.0277 | 0.9259 |
| $\alpha$ | | | | | 1 | 0.9531 | −0.3726 | 0.9012 | 0.6161 |
| $\beta$ | | | | | | 1 | −0.4657 | 0.7577 | 0.6814 |
| $\delta$ | | | | | | | 1 | −0.0798 | −0.9589 |
| $\sum_{i=1}^{n_{pop}} x_i$ | | | | | | | | 1 | 0.3379 |
| Discount | | | | | | | | | 1 |

Figure 46. Arithmetic Metric and Parameter Correlations

It is important to recognize that these measures do not explicitly include the benefits of returning a patient to health.[51] As a result, they are best applied to actions involving illnesses where *existing* treatments are effective. A *new* treatment for an illness that previously was partially or wholly untreatable may increase costs but the benefits must be considered as part of the assessment of value. When considering actions involving an *improved* treatment compared to an existing one that is effective, the measure can be used to make a decision.

* * *

*Conclusion*

Conceptually, healthcare management and wealth management can be thought of as efforts to control a similar process. A quantitative understanding of this is now possible based on the evidence that each is a "diffusion", where the economic objective in healthcare management is on decreasing the rate of growth of expenditures while for wealth management it is on increasing the rate of return on investment. This complements the intuitive understanding of the similarity reflected in the wisdom that "health is wealth".

Policymakers, providers, payors, and patients can influence the healthcare infusion-diffusion process to achieve positive outcomes. Actions to reduce the parameters $\mu$, $\sigma$, $\delta$, $\epsilon$, as well as the fraction of the population who are not ill, will measurably improve value in healthcare.

With the connection established between healthcare and finance, new types of financial instruments designed for healthcare arise as a possibility, which could be issued by the government, corporations, or other entities. To establish a market, they will need to be structured, priced, traded, serviced, and regulated. These instruments could assist, for example, with funding the cost of healthcare, hedging against the risk of illness, increasing the efficiency of the healthcare payment system, measuring healthcare delivery performance against peers, raising capital for healthcare startups, and creating incentives for healthy behaviors such as preventive care. Innovations such as these will contribute to the abundance of healthcare in the future.

*Dedication*

To the NICU kids…their families, and the medical professionals in charge of their care.

*Acknowledgments*

This work was supported by my parents, Joseph and Ida Gardner; my wife Magdalen and son George; my brothers and sisters Tony, Nancy, John, Margo, James, Julie, Grandy, Martha, Maria, Andy, Georgia, and Tasos; my college classmate Skip Bean; and my friends and colleagues. Special thanks are given to the reviewers of the initial draft of the manuscript, Rainer Famulla PhD, Khasha Mohammadi PhD, and Eric Silfen MD, MSHA, MA; the staff of MEPS at the AHRQ who have done the laborious work of assembling a quality data set; as well as to Petter Holme PhD who assisted with placing the manuscript as an e-print on arXiv.org.

The author can be reached at CGardner@i-value.com or by visiting www.i-value.com

APPENDIX A

Figure 47 is the list of data points logarithmically binned $2^i \leq x < 2^{i+1}$ for 2000, 2002, and 2004. The 2004 data point in bin #1 is an outlier. The full data set from 1996-2004 can be found at http://www.meps.ahrq.gov/mepsweb/.

| Bin | x midpoint | 2000 SampleSize | 2000 Population | 2002 SampleSize | 2002 Population | 2004 SampleSize | 2004 Population |
|---|---|---|---|---|---|---|---|
| 1 | 1.41 | 3 | 45053 | 3 | 18232 | 1 | 2169 |
| 2 | 2.83 | 30 | 288136 | 15 | 86998 | 4 | 56524 |
| 3 | 5.66 | 49 | 552856 | 61 | 354179 | 53 | 404746 |
| 4 | 11.31 | 173 | 1824801 | 203 | 1428866 | 114 | 884939 |
| 5 | 22.63 | 295 | 2880061 | 345 | 2119140 | 294 | 2036194 |
| 6 | 45.25 | 907 | 9327880 | 1126 | 7561907 | 984 | 7118483 |
| 7 | 90.51 | 1699 | 18372417 | 2382 | 16043696 | 1953 | 15344526 |
| 8 | 181.02 | 2471 | 28654012 | 3528 | 26201404 | 2819 | 24009426 |
| 9 | 362.04 | 3010 | 35910759 | 4374 | 33998710 | 3723 | 33610563 |
| 10 | 724.08 | 3013 | 37896654 | 4646 | 38463913 | 3805 | 37262194 |
| 11 | 1448.15 | 2802 | 35210769 | 4492 | 38713555 | 3665 | 36843374 |
| 12 | 2896.31 | 2149 | 26846943 | 3981 | 34281025 | 3554 | 36323799 |
| 13 | 5792.62 | 1422 | 17817350 | 2899 | 24374424 | 2807 | 27785269 |
| 14 | 11585.20 | 818 | 10488759 | 1556 | 12971138 | 1599 | 15693053 |
| 15 | 23170.50 | 354 | 4164336 | 697 | 5627045 | 713 | 6909733 |
| 16 | 46341.00 | 144 | 1675706 | 319 | 2467214 | 322 | 3177909 |
| 17 | 92681.90 | 32 | 451903 | 83 | 663460 | 77 | 736108 |
| 18 | 185364.00 | 8 | 109288 | 14 | 91544 | 16 | 218411 |
| 19 | 370728.00 | 1 | 12625 | 4 | 29584 | 3 | 65931 |
| 20 | 741455.00 | 0 | NA | 0 | NA | 1 | 2329 |

**Figure 47. Listing of Data Set (x in units of $)**

APPENDIX B

Figure 48 a,b,c,d is a log-linear {log$_{10}$[x], f(x)} plot of the 4-parameter model for several values of $\mu$, $\sigma$, $\delta$, and $\epsilon$ relative to the best fit for the normalized and combined 2000, 2002, and 2004 data. The baseline parameter values are $\mu$=6.96, $\sigma$=1.62, $\delta$= 5.01x10$^{-7}$ and $\epsilon$= -2.28x10$^{-12}$.

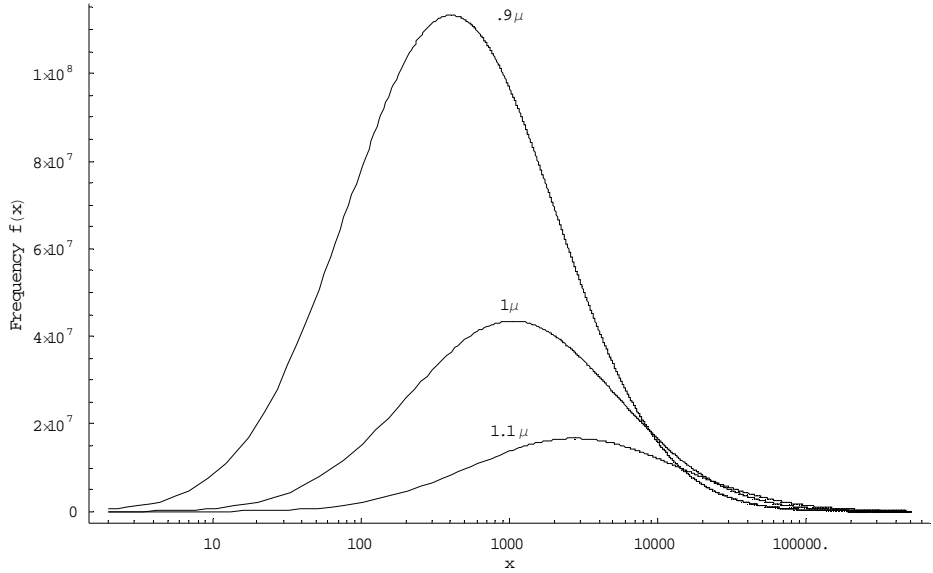

**Figure 48a. Frequency Curve for Three Values of $\mu$**

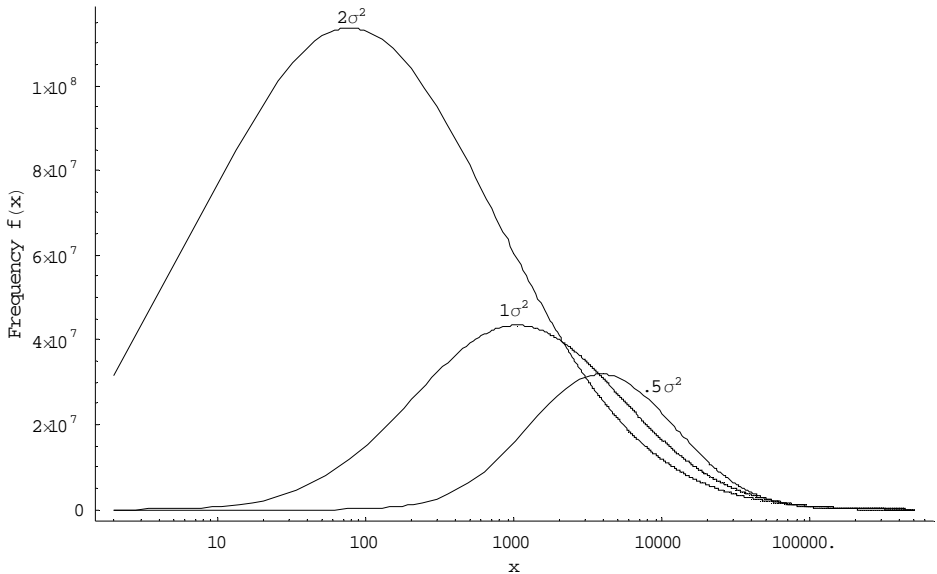

**Figure 48b. Frequency Curve for Three Values of $\sigma^2$**

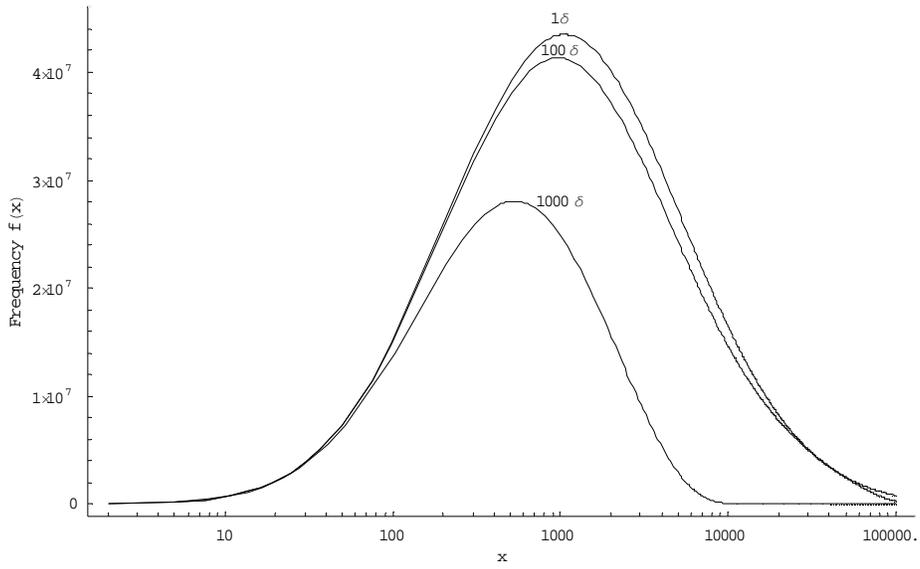

**Figure 48c. Frequency Curve for Three Values of $\delta$**

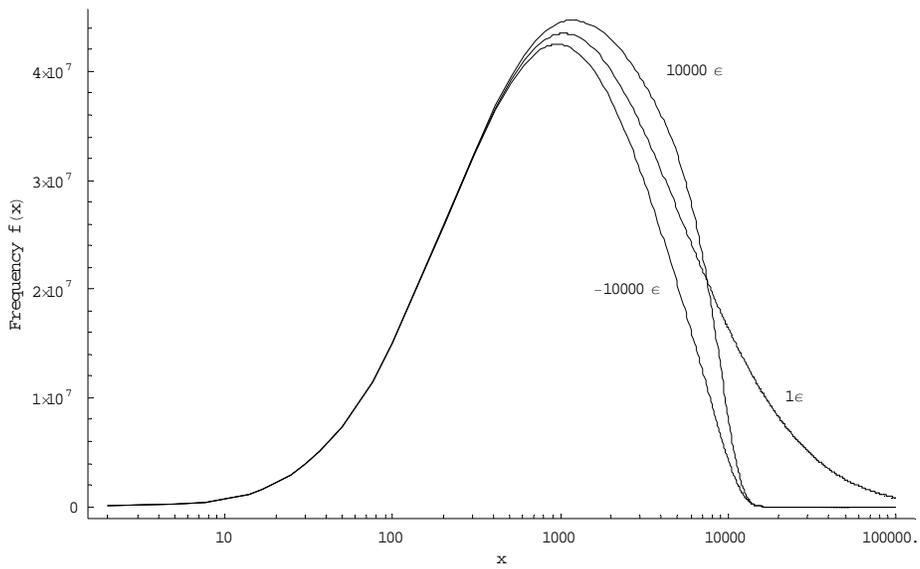

**Figure 48d. Frequency Curve for Three Values of $\epsilon$**

[1] F. Dominici, G. Parmigiani, S. Venturi; *Gamma Shape Mixtures for Heavy-Tailed Distributions*; Johns Hopkins University Department of Biostatistics Working Paper #124 2006

[2] D. Feenberg and J. Skinner; *The Risk and Duration of Catastrophic Healthcare Expenditures*; The Review of Economics and Statistics; 76(4) 633-647 1994

[3] J. Rust and C. Phelan; *How Social Security and Medicare Affect Retirement Behavior in a World of Incomplete Markets*; Econometrica; 65(4) 781-831

[4] E. French and J. Jones; *On the Distribution and Dynamics of Health Care Costs*; The Federal Reserve Bank of Chicago; 2004

[5] J. Aitchison and J. Brown; *The Lognormal Distribution: with Special Reference to its Uses in Economics*; Cambridge University Press 1969

[6] W. Eaton and G. Whitmore; *Length of Stay as a Stochastic Process: A General Approach and Application to Hospitalization for Schizophrenia*; Journal of Mathematical Sociology; 5: 273-292 1977

[7] M. Von Korff; *A Statistical Model of the Duration of Mental Hospitalization: The Mixed Exponential Distribution*; Journal of Mathematical Sociology; 6: 169-175 1979

[8] S. Schalkowsky and R. Wiederkehr; *Log-Normal Model for Microbial Survival in Heat Sterilization*, NASA Topical Report TR-015; October 1966

[9] V. Plerou, P. Gopikrishnan, B. Rosenow, L. Amaral, H. Stanley; *Econophysics: Financial Time Series from a Statistical Physics Point of View* Physica A279 (2000) 443-456

[10] D.R. Cox and H.D. Miller; *The Theory of Stochastic Processes*; John Wiley and Sons 1966

[11] A. Dixit and R. Pindyck; *Investment Under Uncertainty*; Princeton University Press 1994

[12] In science, it is called geometric (or exponential) Brownian motion with drift (after the botanist Robert Brown), or a Wiener process with drift (after the mathematician Norbert Weiner). It is used as a model for stock prices (by the economist Paul Samuelson), as the basis for stock option pricing (the Black-Scholes model), and the description of phenomenon such as drift, diffusion, fluctuations, and the dissipation of energy of particles in suspension (Einstein, von Smoluchowski, Langevin, Onsager, et. al).

[13] W. Press, S. Teukolsky, W. Vetterling, B. Flannery; *Numerical Recipes in C: The Art of Scientific Computing*; Cambridge University Press 2002

[14] The analysis can be reproduced with a laptop PC, a symbolic mathematics software package such as Mathematica, and US government supplied healthcare expenditure data that is publicly available on the Internet

[15] J.W. Cohen; *Design and methods of the Medical Expenditure Panel Survey Household Component*; Agency for Health Care Policy and Research; 1997 MEPS Methodology Report No.1. AHCPR Pub. No. 97-0026

[16] *Online Workbook: MEPS Survey Overview*; pg 7 Agency for Healthcare Research and Quality, US Department of Health and Human Services 2007

[17] William W. Yu; *An Approximation of Skewed Healthcare Expenditure Distribution Using a Mixture Model;* Agency for Healthcare Research and Quality; Proceedings of the Survey Research Methods Section, American Statistical Association 2006

[18] This "right-hand" binning $2^i \leq x < 2^{i+1}$ is compatible with the focus on expenditures ≥$1 and exhibits some differences with "left-hand" binning $2^i < x \leq 2^{i+1}$. Most notably, the choice of bin boundaries results in an outlier data point in the first interval i=0 for 2004.

[19] M. H. R. Stanley, S.V. Buldyrev, S. Havhn, R. N. Mantegna, M. A. Salinger, H. E. Stanley; *Zipf Plots and the Size Distribution of Firms*; Economics Letters 49 (1995) 453-457

[20] This is the "Law of Proportionate Effect".

[21] This derivation is a variation of that in J. Kapteyn; *Skew Frequency Curves in Biology and Statistics*; Astronomical Laboratory, Noordhoff, Groningen 1903

[22] J. Aitchison and J. Brown; pg 8 *op. cit.*

[23] F. Dominici, G. Parmigiani, S. Venturi; *op. cit.*

[24] E. French and J. Jones; *op. cit.*

[25] J. Aitchison and J. Brown; pg 8 *op. cit.* This transformation is discussed in N. Duan, W. Manning, C. Morris, J. Newhouse *A Comparison of Alternative Models for the Demand for Medical Care* Rand 1982 R-2754-HHS

[26] The proof of this relation is

$$x\, p(x) = \frac{C_x}{C_{\ln[x]}}\, p(\ln[x])$$

taking the expected value of both sides gives

$$e^{\mu + \frac{\sigma^2}{2}} = \frac{C_x}{C_{\ln[x]}}$$

[27] These tests reveal the cautionary fact that the formula for ln[p(x)] is substantially better than nonlinear curve fit algorithms used for predicting the parameters of the classic lognormal curve fit, which hopefully is a rare occurrence.

[28] The best fit for the classic lognormal conforms to the relation between ln[p(x)] and ln[p(ln[x])] that generates a shift in the constant term

$$\ln[C_x] + \ln\left[\frac{n_{pop}}{\sqrt{2\pi\sigma^2}}\right] = \ln[C_{\ln[x]}] + \ln\left[\frac{n_{pop}}{\sqrt{2\pi\sigma^2}}\right] + \mu + \frac{\sigma^2}{2} = 17.6705 + 6.8491 + \frac{1.5783^2}{2} = 25.765$$

[29] D. Bates and D. Watts; *Nonlinear Regression Analysis and Its Applications*, John Wiley & Sons 1988 p108

[30] A nationwide random sample could probably be developed from existing data with access and effort. Large private healthcare insurance companies, for example, might be able to assemble a random sample from files on payment histories and copayment policies for the subpopulations they underwrite. This would have to be combined with random samples for missing subpopulations such as those publicly insured by Medicare/Medicaid.

[31] A nested model is a model that is a special case of another. In our case the classic lognormal is nested in the 3-parameter lognormal with infusion model, as can be seen when $\delta=0$; and the 3-parameter lognormal with infusion model is nested in the 4-parameter model as can be seen when $\epsilon=0$.

[32] The KS- extra sum of squares procedure is necessary because the straightforward method of simply calculating the K-S statistic for each model is not possible since we lack expressions for the CDF of the 3 and 4 parameter lognormal with infusion models

[33] Some publications use the Kolmogorov-Smirnov probability function to calculate this significance. However, "if you estimate any parameters from a data set…then the distribution of the KS statistic…is no longer given by (the Kolmogorov-Smirnov probability function)…you will have to determine the distribution yourself, e.g., by Monte Carlo methods" p627 W. Press, S. Teukolsky, W. Vetterling, B. Flannery; *op. cit.*

[34] Normally a low p-value is considered good, because the null hypothesis assumes that a model other than the one that we are attempting to verify is correct. A low value then rules this out. Here we use the p-value to confirm the hypothesis that the model we are attempting to verify is correct. So a high p-value is good.

[35] Private communication with Eric Silfen, MD, MSHA, MA; December 2007

[36] A. Clauset, C. Shalizi, and M. Newman; *Power Law Distributions in Empirical Data*, http://arxiv.org/abs/0706.1062v1 June 2007

[37] M. Goldstein, S. Morris, and G. Yen; *Problems with Fitting to the Power-Law Distribution*, arXiv:cond-mat/0402322 v3 13 Aug 2004 and A. Clauset, C. Shalizi, and M. Newman; *op. cit.*

[38] For a discussion of the comparison of a lognormal to a power law distribution and the non-nested likelihood ratio test, see A. Clauset, C. Shalizi, and M. Newman; *op. cit.*

[39] J. Rust and C. Phelan; How Social Security and Medicare Affect Retirement Behavior in a World of Incomplete Markets; Econometrica 65(4), 781-831 1997

[40] B. G. Druss, S. C. Marcus, M. Olfson, and H. A, Pincus; *The Most Expensive Medical Conditions in America,* Health Affairs 21:44, 105-111, 2002. The power law pattern can be seen in Exhibit 2 of this article.

[41] S. Roman; *Introduction to the Mathematics of Finance*, Springer Verlag pg 240

[42] This can be seen as follows:

$$\ln[x(t)] - \ln[x_0] = \delta x + N[\mu_* t + \mu_o, \sigma^2(t)]$$

$$x(t) = x_0\, e^{\delta x + N[\mu_* t + \mu_o, \sigma^2(t)]} = x_0\, e^{\mu_* t}\, e^{\delta x + N[\mu_o, \sigma^2(t)]}$$

Therefore $\mu_*$ is the continuously compounded annual growth rate and 1-Exp[$\mu_*$] is the cumulative annual growth rate of expenditures.

[43] J. Grossman and P. Ginsburg; *As the Health Insurance Underwriting Cycle Turns: What Next?*, Health Affairs V23:6 Nov-Dec 2004 and A. Rosenblatt; *The Underwriting Cycle: The Rule of Six (Perspective)*; Ibid

[44] To compare the mean and variance of healthcare expenditures with GDP, the data has been detrended with a Hodrick-Prescott filter used for analyzing business cycles with $\lambda=100$ for annual data. The comparison is relatively insensitive to $\lambda$, since we only care about the general form of the cycles. The results represent deviations from the trend which reveals the underlying cyclic pattern. The business cycle is measured by the deviation in the logarithm of the Gross Domestic Product from trend. The healthcare expenditure cycle is measured by the deviation in the mean and variance of ln[x] from trend. The healthcare expenditure and GDP data are in nominal terms.

[45] For example,

$$t < \sigma^2(t) > = \int_0^t \sigma^2(t) \, dt$$

$$\sigma^2(t) = \frac{d[t < \sigma^2(t) >]}{dt}$$

$$\sigma^2(t) = \frac{d[t(2.772 + 0.005\,t)]}{dt} = 2.772 + 2*0.005\,t$$

[46] This can be seen as follows:

$$x(t + \Delta t) = x_0 \, e^{\delta x + N[\mu_*(t+\Delta t), \sigma_*^2(t+\Delta t)] + N[\mu_o, \sigma_o^2]}$$

$$x(t + \Delta t) = x(t) \, e^{N[\mu_* \Delta t, \sigma_*^2 \Delta t]}$$

Setting $\Delta t = 1$ year, and noting that whenever we want to get the standard deviation over a longer time span $T=k\Delta t$ we must multiply it by the square root of k giving $\sigma_*\sqrt{k}$

$$x(t + \Delta t) = x(t) \, e^{N[\mu_*, \sigma_*^2]}$$

Therefore we can conclude that $\sigma_*$ is the standard deviation (i.e., volatility) of the continuously compounded annual rate of growth of expenditures $\mu_*$

[47] The logarithmic form of the equation is *arithmetic* Brownian motion with drift. The exponential form is *geometric* Brownian motion with drift.

[48] When the discount is 10% the value is .95 for the ratio

$$\frac{1 - e^{-\Delta \mu}}{\Delta \mu}$$

[49] A decrease is "good" while an increase is "bad"

[50] The information factor is in the same category as the "information ratio" which is defined as the mean over the standard deviation, with no infusion term. It is similar to the "Sharpe Ratio" in finance.

[51] Though benefits are not explicitly included, they are implicitly included to the extent that healthcare costs are proportional to the time spent in treatment. The benefits are then accounted for as the time *not* spent in treatment.

___